\documentclass[aps,article,showpacs,showkeys]{revtex4}%
\usepackage{amsmath,bm}
\usepackage{amsmath}
\usepackage{amsfonts}
\usepackage{amssymb}
\usepackage{graphicx}%
\input{psfig.sty}

\def\beq{\begin{eqnarray}}
\def\eeq{\end{eqnarray}}

\def\0{\mbox{\boldmath$\displaystyle\mathbf{0}$}}

\def\1{\mbox{\boldmath$\displaystyle\mathbf{1}$}}

\def\h00h{\mbox{\boldmath$\displaystyle\mathbf{(1/2,0)\oplus(0,1/2)}$}}

\begin{document}
%\date{\today }

\title{Spin-$\frac{3}{2}$ Beyond the Rarita-Schwinger Framework.}
\author{Mauro Napsuciale$^{*}$, Mariana Kirchbach$^{**}$, Sim\'{o}n Rodriguez$^{**}$}
\address{$^{*}$ Insituto de Fisica, Univ.\ de Guanajuato, Lomas del Bosque 103,
Fracc.\ Lomas del Campestre, 37150 Le\'{o}n, Guanajuato, M\'{e}xico. \\
$^{**}$ Instituto de Fisica, Univ. Aut. de San Luis Potos\'{\i},
 Av. Manuel Nava 6, Zona Universitaria,
 78290,  San Luis Potos\'{\i}, SLP, M\'{e}xico.}

 \begin{abstract}
We employ the  two independent Casimir operators of the 
Poincar\'e  group, the squared four--momentum, $p^{2}$, and the 
squared Pauli-Lubanski vector, $\mathcal{W}^{2}$, 
in the construction of a covariant mass-$m$, and 
spin-$\frac{3}{2}$ projector in 
the four--vector--spinor, $\psi_{\mu}$. 
This projector provides the basis for 
the construction of an interacting  Lagrangian that 
describes a causally propagating  spin-$\frac{3}{2}$ particle 
coupled to the electromagnetic field
by a gyromagnetic ratio of $g_{\frac{3}{2}}=2$. 
\end{abstract}

\maketitle

\section{Introduction.}

High spin particles occupy an important place in theoretical physics. For the
first time they were observed as resonant excitations in pion-nucleon
scattering. The Particle Data Group \cite{PART} lists more than thirty
non-strange baryon resonances with spins ranging from $\frac{3}{2}$ to
$\frac{15}{2}$, and more than twenty strange ones with spins from $\frac{3}%
{2}$ to $\frac{9}{2}$. Baryon resonances have been extensively investigated in
the past among others at the former Los Alamos Meson Physics Facility (LAMPF), 
and at present their study continues at the Thomas Jefferson National 
Accelerator
Facility (TJNAF) \cite{Burk_Lee}. Such particles are of high relevance in the
description of photo- and electro-pion production off proton, where they
appear as intermediate states, studies to which the Mainz Microtron (MAMI)
devotes itself since many years \cite{MAMI}. Search for high-spin solutions to
the QCD Lagrangian has been recently reported by the Lattice collaboration in
Ref.~\cite{32llatt}. Moreover, also the twistor formalism has been employed in
the construction of high spin fields \cite{32_twistor}. Integer high-spin
meson resonances with spins ranging from 0 to 6 can have importance in various
processes revealing the fundamental features of QED at high energies such like
pair production \cite{Kruglov}. However, the most attractive high-spin fields
appear in proposals for physics beyond the standard model which invoke
supersymmetry \cite{Kaku} and contain gauge fields of fractional spins such as
the gravitino-- the supersymmetric partner of the ordinary spin-$2$ graviton.
Supersymmetric theories open the venue to the production of fundamental 
spin-$\frac{3}{2}$ particles at early stages of the universe, 
whose understanding
can play an important role in its evolution \cite{Verdi}.

The description of high spins takes its origin from Refs.~\cite{FiePau},
\cite{RS}, \cite{Weinbrg65} which suggest to consider any fractional spin-$s$
as the highest spin in the traceless and
totally symmetric rank-$(s-\frac{1}{2})$ Lorentz
tensor with Dirac spinor components, $\psi_{\mu_{1}...\mu_{s-\frac{1}{2}}}$.
{}For spin-$\frac{3}{2}$ one has to consider the four-vector--spinor, $\psi
_{\mu}$,
\begin{equation}
\psi_{\mu}=A_{\mu}\otimes\psi\simeq\left(  \frac{1}{2},\frac{1}{2}\right)
\otimes\left[  \left(  \frac{1}{2},0\right)  \oplus\left(  \frac{1}%
{2},0\right)  \right]  \,, \label{Psi_mu}%
\end{equation}
the direct product between the four vector, $A_{\mu}$, and the Dirac spinor,
$\psi$, and solve the system of three linear (in the momenta) equations
\begin{eqnarray}
(\not p  -m)\psi_{\mu}  &  =&0\,,\label{RS_Dirac}\\
\gamma^{\mu}\psi_{\mu}  &  =&0\,,\label{RS_Proca}\\
p^{\mu}\psi_{\mu}  &  =& 0\,, \label{RS_What}%
\end{eqnarray}
known as the Rarita-Schwinger (RS) framework. 
Next one designs~\cite{MC} the most general 
family of Lagrangians depending on the undetermined 
parameter $(A)$, with the aim to reproduce 
Eqs.~(\ref{RS_Dirac})-(\ref{RS_What}).
 The Lagrangians obtained this way  read
\begin{eqnarray}
\mathcal{L}^{(RS)}(A) & =&\overline{\psi}^{\mu}
\left(  p_{\alpha}\Gamma_{\mu\quad\nu}^{\quad\alpha}(A)- m~B_{\mu
\nu}(A)\right) \psi^{\nu}\, ,
\end{eqnarray}
where
\begin{eqnarray}
\Gamma_{\mu\quad\nu}^{\quad\alpha}(A)&=&g_{\mu\nu}\gamma_{\alpha}
+A(\gamma_{\mu}g^{\alpha}_{\nu}+g_{\mu}^{\alpha}\gamma_{\nu})
+B\gamma_{\mu} \gamma^{\alpha}\gamma_{\nu},  \nonumber \\
B_{\mu\nu}(A)&=&g_{\mu\nu}-C\gamma_{\mu}\gamma\nu ,
 \label{Teil_2}\\ 
A\neq\-\frac{1}{2}, \qquad B&\equiv&\frac{3}{2}A^{2}+A+\frac{1}{2}, 
\qquad C\equiv3A^{2}+3A+1. \nonumber %
\end{eqnarray}
The case $A=-\frac{1}{3}$ corresponds to the Lagrangian 
originally proposed in \cite{RS}.  Another value widely used in 
the literature is $A=-1$ in which case the Lagrangian simplifies to 
\begin{equation}
\mathcal{L}^{(RS)}(A=-1) =\overline{\psi}^{\mu}
\left(  p_{\alpha}\epsilon_{\mu\quad\nu\rho}^{\quad\alpha}
\gamma^{5}\gamma^{\rho}
- im~\sigma_{\mu\nu}\right) \psi^{\nu}. \label{super}
\end{equation}
If we define 
\begin{equation}
K_{\mu\nu}(A)= p_{\alpha}\Gamma_{\mu\quad\nu}^{\quad\alpha}(A)- 
m~B_{\mu\nu}(A),
\end{equation}
the above Lagrangian factorize as
\begin{equation}
\mathcal{L}^{(RS)}(A)  =\overline{\psi}^{\mu}~
R_{\mu\rho}\left(\frac{A}{2}\right)~
K^{\rho\sigma}(0)R_{\sigma\nu}\left(\frac{A}{2}\right) \psi^{\nu}\, ,
\end{equation}
where
\begin{equation}
R_{\mu\rho}(w)\equiv g_{\mu\rho}+w \gamma_{\mu}\gamma_{\rho}.\label{rotation}
\end{equation}
This factorization can be used to show that the Lagrangian is 
invariant under the point transformations 
\begin{equation}
\psi_{\mu}\to\psi^{\prime}_{\mu}=R_{\mu\nu}(w)\psi^{\nu},\quad 
A\to \frac{A-2w}{1+4w}. \label{point}
\end{equation}

Over the years, Eqs.~(\ref{RS_Dirac})-(\ref{RS_What}) have been widely applied
in hadron physics to the description of predominantly the $\Delta(1232)$-- and
occasionally the $D_{13}(1520)$ resonances and their contributions to various
processes. Recent applications of the Rarita-Schwinger spin-$\frac{3}{2}$
description to calculations of light-hadron properties along the line of
Chiral Perturbation Theory can be found in Refs.~\cite{32_CHPT,Hemmert}.

\noindent The freedom represented by the parameter $A$ reflects
 invariance under 
"rotations" mixing the two spin-$\frac{1}{2}^{+}$ and $\frac{1}{2}^{-}$ 
sectors residing in the RS representation space besides 
spin-$\frac{3}{2}$ \cite{Benm,NL}. 
It can be shown  \cite{KOS} that the elements of the $S$ matrix 
do not depend on the parameter $A$. Yet, this symmetry, when 
implemented into the interacting theory, introduces ambiguities 
represented by free parameters, the so called "off-shell" 
parameters \cite{Benm,NL,Gabriel,Mohamed}. This is not
to remain the only disadvantage of the RS framework. A detailed study of
Eqs.~(\ref{RS_Dirac}), (\ref{RS_Proca}), and (\ref{RS_What}) revealed that
the Rarita-Schwinger framework suffers some more fundamental weaknesses. The
quantization of the interacting spin-$\frac{3}{2}$ field turned out to be
inconsistent with Lorentz covariance, an observation reported by Johnson and
Sudarshan in Ref.~\cite{sudarshan1}. Furthermore, the wave fronts of the
\underline{classical} solutions of the Rarita-Schwinger spin-$\frac{3}{2}$
equations were shown to suffer acausal propagation within the electromagnetic
environment, an observation due to Velo and Zwanziger \cite{VZ1,VZ2}. 
This is an old problem and several remedies have been suggested over the years 
\cite{Pascags,RaSierra,SIK}. In \cite{RaSierra}, 
it was shown that the standard Rarita-Schwinger description allows
to avoid the Velo-Zwanziger problem only to the cost of 
propagating simultaneously twelve degrees of freedom
associated with  spin-$\frac{1}{2}$, and spin-$\frac{3}{2}$. 
The more recent reference \cite{SIK} suggests two
new wave equations in $\psi_{\mu}$, one of which is linear
local, and the other, quadratic and  non-local.
The local equation propagates causally all sixteen degrees 
of freedom in $\psi_{\mu}$ associated with the spin-cascade
$\left( \frac{1}{2}^+,\frac{1}{2}^-, \frac{3}{2}^-\right)$, 
but weighted with three different masses. The non-local equation 
propagates causally the twelve degrees of freedom
corresponding to spin-$\frac{1}{2}^-$ and $\frac{3}{2}^-$
treated as mass-degenerate. The latter results indicate
that the description of a causal single spin-$\frac{3}{2}$ 
propagation is beyond the reach of the Rarita-Schwinger 
framework.

It is the goal of the present work to construct a single-spin-$\frac{3}{2}$
Lagrangian and associated wave equation such that
the wave fronts of its solutions propagate causally
within an electromagnetic environment and the spin-$\frac{3}{2}$ particle
is coupled to the electromagnetic field through a gyromagnetic
ratio of $g_{\frac{3}{2}}=2$ as required by unitarity in the 
ultrarelativistic limit~\cite{Weinberg,Djukanovic}.
The Lagrangian in question is entirely based upon the 
Poincar\'e group generators in  $\psi_\mu$
and the magnetic coupling is identified in a fully 
covariant fashion. Compared to this, within the Rarita-Schwinger framework the 
gyromagnetic factor is extracted at the non-relativistic level 
\cite{MC,Weda,Nozawa} or from calculating  pion-nucleon
bremsstrahlung and a subsequent comparison to low energy theorems
\cite{Psct_EM}.

The paper is organized as follows.
In the next  Section we outline the general procedure of pinning down an 
invariant subspace of mass-$m$ and spin-$s$ on the example of a 
generic Lorentz group
representation containing two Poincar\'{e} invariant spin-sectors, for
simplicity. There, we further present the associated second order (in the
momenta) equation of free motion. As a consistency check for our suggested
formalism we re-derive there the Proca equation in applying the procedure to
$\left(  \frac{1}{2},\frac{1}{2}\right)  $.
In Section 3 we  apply above procedure to the four-vector--spinor and
derive the corresponding equation of motion, the associated
Lagrangian, and the respective propagator.
Section 4 is devoted to the symmetries of the suggested Lagrangian in the
massless limit and its relation to "rotations" within the spin-$\frac{1}{2}$ 
sector. 
In Section 5 we introduce electromagnetic interactions.
The paper closes with a brief summary and has one Appendix.

\bigskip

\section{Particle dynamics and Poincar\'e group invariants.}

\subsection{The Casimir operators $p^{2}$ and $\mathcal{W}^{2}$ and their
invariant vector spaces.}

In the present work we aim to identify spin-$\frac{3}{2}$ directly and in a
covariant fashion according to the conventional understanding of a particle as
an invariant vector space of the two Casimir invariants of the 
Poincar\'{e} group, the
first being the squared four-momentum, $p^{2}$, and the second, the squared
Pauli-Lubanski vector, $\mathcal{W}^{2}$. Accordingly, the 
corresponding states must be labeled by the
eigenvalues of these operators (see Ref.~\cite{ED} for details),
\begin{eqnarray}
p^{2}\Psi^{(m,s)}  &  =&m^{2}\Psi^{(m,s)}\,,\label{PGR_KG}\\
\mathcal{W}^{2}\Psi^{(m,s)}  &  =&-p^{2}s(s+1)\Psi^{(m,s)}\,. \label{PGR_LBL}%
\end{eqnarray}
Here $m$ stands for the mass, while $\Psi^{(m,s)}$ denotes a generic
Poincar\'{e} group representation of mass $m$ and rest-frame spin $s$. 
Equation (\ref{PGR_KG}) is the Klein-Gordon equation that fixes the mass of
the states, while Eq.~(\ref{PGR_LBL}) fixes the spin. As already mentioned in
the introduction, the mass--shell condition as reflected by the Klein-Gordon
equation has been of wide use in the formulation of free-particle Lagrangians,
not so the spin condition.  We shall formulate a new Lagrangian formalism 
that incorporates Eqs.~(\ref{PGR_KG}), (\ref{PGR_LBL}) on equal footing and 
obtain a single condition on the field that encodes both 
the mass-shell-- , and spin-conditions.
In so doing we first have to resolve the notorious problem of
non-coincidence between Poincar\'{e} and Lorentz group labels that occurs for
all representations beyond $(s,0)\oplus(0,s)$.

Indeed, Lorentz representations are labeled by the so called
left $(s_{L})$, and right-handed, $(s_{R})$, "spins", the respective
eigenvalues to $\vec{S\,}_{L\,}^{2}=\frac{1}{4}\left(  \vec{J\,}+i\vec
{K\,}\right)  ^{2}$, and $\vec{S\,}_{R}^{2}=\frac{1}{4}\left(  \vec{J\,}%
-\vec{iK\,}\right)  ^{2}$, where $\vec{J}$, and $\vec{K}$ represent the
generators of rotations and boost in the basis of interest. The Poincar\'{e}
$s$--label enters the Lorentz representation,
$\Psi^{\left( m,(s_{L},s_{R})\right) }$, via 
$s=|s_{L}-s_{R}|,|s_{L}%
-s_{R}|+1,...,(s_{L}+s_{R})$, which causes reducibility of 
$\Psi^{\left(  m,\left( 
s_{L},s_{R}
\right)\right)  }$ 
into the following Poincar\'{e} invariant subspaces:
\begin{equation}
\Psi^{\left(m,\left(  s_{L},s_{R}\right)\right)  }\longrightarrow
\Psi^{\left( m,|s_{L}-s_{R}|\right) }\oplus
\Psi^{\left( m,|s_{L}-s_{R}|+1\right) }
\oplus...\oplus\Psi^{\left(m,(s_{L}%
+s_{R})\right)  }\,. \label{LP_red}%
\end{equation}
The problem one is facing  now is the covariant tracking of the 
sector of interest.
In the next subsection we formulate our procedure for the covariant
tracking of the highest spin-$\frac{3}{2} $ of mass- $m$ in the vector-spinor 
representation $\psi_{\mu}$.

\subsection{Covariant mass-$m$ and spin-$s$ tracking procedure.}

\subsubsection{The general case.}
We begin by noticing that in general the generators of the Poincar\'{e} group
are marked by external space-time ( Lorentz) indices (see Appendix) and which
we denote by small Greek letters $\mu,\nu,\lambda,\rho$, etc. next to
representation specific indices (denoted by capital Latin letters $A,B,C,...$
etc.) At times, like for example in $\psi_{\mu}$, it may be
possible and useful to separate the capital Latin letter
indices into Lorentz- and spinorial parts. Therefore, the most general form of
the Pauli-Lubanski vector operator is given by%
\begin{equation}
(\mathcal{W}_{\lambda})_{AC}=\frac{1}{2}\epsilon_{\lambda\rho\sigma\mu
}(M^{\rho\sigma})_{AC}p^{\mu}. \label{W}%
\end{equation}
Its squared is then written as
\begin{equation}
(\mathcal{W}_{\lambda}\mathcal{W}^{\lambda})_{AB}=\frac{1}{4}\epsilon
_{\lambda\rho\sigma\mu}(M^{\rho\sigma})_{AC}p^{\mu}\epsilon_{~\tau\xi\nu
}^{\lambda}(M^{\tau\xi})_{CB}p^{\nu}
\equiv T_{AB\mu\nu}p^{\mu}p^{\nu}.
\label{W2}%
\end{equation}
Notice that $T_{AB\mu\nu}$  is momentum independent.

\begin{quote}
Our pursued spin-tracking strategy will be the construction of covariant
projectors onto the Poincar\'e invariant $\Psi^{( m, s) }$ sectors of the
Lorentz representation of interest. Below we illustrate this procedure for the
simplest case of a generic Lorentz representation having only two Poincar\'e
invariant subspaces with spins differing by one unit. We denote the 
maximal and minimal spins by $s$ and $(s-1)$, 
respectively.
\end{quote}

The covariant mass-$m$ and spin-$s$ tracking procedure can be outlined as 
follows.
Construct the Poincar\'{e} covariant mass-$m$--spin-$s$, and 
mass-$m$--spin-$(s-1)$ projectors as
\begin{eqnarray}
\mathcal{P}^{(m;s)}(p)  &  =&-\frac{1}{2s}\left(  \frac{\mathcal{W}^{2}}{m^{2}%
}+s(s-1)\frac{p^{2}}{m^{2}}\mathbf{1}_{n\times n}\right),\label{maxs_pr}\\
\mathcal{P}^{(m;s-1)}(p)  &  =&\frac{1}{2s}\left(  
\frac{\mathcal{W}^{2}}{m^{2}%
}+s(s+1)\frac{p^{2}}{m^{2}}\mathbf{1}_{n\times n}\right), \label{mins_pr}%
\end{eqnarray}
where $n$ stands for the dimensionality of the representation of interest. We
must remark that these operators are projectors over well defined spins
whenever the particles are on mass shell. Indeed, using the basis of
eigenstates of $\mathcal{W}^{2}$ it can be easily shown that on mass shell
they satisfy the following relationships:
\begin{eqnarray}
\left[  \mathcal{P}^{(m;s)}(p)\right]  ^{2}&=&\mathcal{P}^{(m;s)}(p), \qquad
\left[  \mathcal{P}^{(m;s-1))}(p)\right]  ^{2}=\mathcal{P}^{(m;s-1)}%
(p)\,,\label{cov-prj} \nonumber \\
\mathcal{P}^{(m;s)}(p)\mathcal{P}^{(m;s-1)}(p)&=&0,\qquad \mathcal{P}^{(m;s)}(p) +\mathcal{P}^{(m;s-1)}(p)=\mathbf{1}_{n\times n}\,.
\end{eqnarray}
The important point here is that in the general case, imposing the condition
\begin{equation}
\mathcal{P}^{(m;s)}(p)\Psi^{(m,s)\ }=\Psi^{(m,s)\ }, \label{prj_eq}%
\end{equation}
will simultaneously track down the desired spin-$s$, nullify spin-$(s-1)$, and
incorporate the mass shell condition%
\begin{eqnarray}
\mathcal{P}^{(m;s)}(p)\Psi^{(m,(s-1)) \ }  &  =&0\,,\label{cov_pr}\\
(p^{2}-m^{2})\Psi^{(m,s)\ }  &  =&0.
\end{eqnarray}
Thus our projectors simultaneously track down well defined mass-$m$-- and well
defined spin-$\ s$ or spin-$(s-1)$ eigenspaces. In its most general form,
Eq.~(\ref{prj_eq}) can be written as
\begin{equation}
\left[  -\Gamma_{AB\mu\nu}p^{\mu}p^{\nu}+m^{2}\delta_{AB}\right]  \Psi
_{B}^{(m,s)}=0, \label{Platino0}%
\end{equation}
where%
\begin{equation}
\Gamma_{AB\mu\nu}=-\frac{1}{2s}\left(  T_{AB\mu\nu}+s(s-1)\delta_{AB}%
\ g_{\mu\nu}\right)  \,,\label{Gamma_tensor}%
\end{equation}
with $T_{AB\mu\nu}$ defined in  Eq.~(\ref{W2}) from above.
Compared to Ref.~\cite{MK03}, the $\mathcal{P}^{(m;s)}(p)$ projectors 
contain the
additional factor of $p^{2}/m^{2}$ in front of $s(s-1)\mathbf{1}_{n\times n}$
which is indispensable for fixing correctly the mass of the tracked state as
visible through Eq.~(\ref{cov_prj_fct}) below.

\subsubsection{The spin-1 case}

The most important examples for applications of the covariant spin-tracking
procedure are the four-vector, $A_{\mu}$, and the four-vector spinor,
$\psi_{\mu}$. In the former case, using the explicit form for $W^{2}$ in
Eq.~(\ref{W2_hh}) from the Appendix we obtain 
\begin{equation}
\Gamma^{P}_{\alpha\beta\mu\nu}=g_{\alpha\beta}g_{\mu\nu}-
g_{\alpha\nu}g_{\beta\mu},
\end{equation}
and Eqs.~(\ref{cov_pr}),(\ref{Platino0}) yield just the Proca equation
\begin{equation}
\left[ (-p^{2}+m^{2}) g_{\mu\nu}+p_{\mu}p_{\nu} \right] 
A^{\nu}=0, \label{proca_pak}%
\end{equation}
which can be derived from the following Lagrangian:
\begin{eqnarray}
\mathcal{L}_{P}&=&-\frac{1}{2}(\partial^{\mu}A^{\alpha})\Gamma^{P}_{\alpha
\beta\mu\nu}\partial^{\nu}A^{\beta}+\frac{m^{2}}{2}A^{\alpha }A_{\alpha} 
\nonumber \\
&=& -\frac{1}{4}F^{\mu\nu }F_{\mu\nu}+\frac{m^{2}}{2}A^{\alpha }A_{\alpha}.
\label{NKlag_proca}%
\end{eqnarray}
It is quite instructive to rewrite Eq.~(\ref{proca_pak}) in terms of the
spin-$1$ and spin-$0$ projectors in $A_{\mu}$, in turn denoted 
by ${\mbox{\bf P}}_{\mu\nu}^{(1)}$, and ${\mbox{\bf P}}_{\mu\nu}^{(0)}$, 
and defined as%
\begin{equation}
{\mbox{\bf P}}_{\mu\nu}^{(1)}=\frac{(W^{2})_{\mu\nu}}{-2p^{2}}=g_{\mu\nu}%
-\frac{p_{\mu}p_{\nu}}{p^{2}}\,,\quad{\mbox{\bf P}}_{\mu\nu}^{(0)}=
\frac{p_{\mu }p_{\nu}}{p^{2}}\,. \label{proj1}
\end{equation}
In so doing one finds
\begin{equation}
{\mathcal P}^{(m,1)}_{\mu\nu}(p) A^{\nu}=
\frac{p^{2}}{m^{2}}{\mbox{\bf P}}_{\mu\nu}^{(1)}A^{\nu}=A_{\mu}%
\,,\label{revelation}%
\end{equation}
an equation which reveals the Poincar\'{e}  invariant projector
$\mathcal{P}^{(m,1)}(p)$ as the direct product of the 
mass-$m$ and spin-$1$ projectors. The
inverse to equation (\ref{proca_pak}) provides the Proca propagator as%
\begin{equation}
\Pi_{\mu\nu}^{\mbox{\footnotesize Proca}}
= \frac{\Delta_{\mu\nu}^{ \mbox{\footnotesize Proca }}}
{\left(  p^{2}-m^{2}+i\varepsilon
\right) }
\, , \label{Proca-Prop}%
\end{equation}
where
\begin{equation}
\Delta_{\mu\nu}^{\mbox{\footnotesize Proca }}=
-g_{\mu\nu}+\frac{p^{\mu}p^{\nu}}{m^{2}}=-{\mbox{\bf P}}%
_{\mu\nu}^{(1)}+\frac{p^{2}-m^{2}}{m^{2}}{\mbox{\bf P}}_{\mu\nu}^{\left(
0\right)  }. \label{prop_1}%
\end{equation}
Throughout this paper, propagators are given in momentum space,
hence momentum operators like $p^{\mu}\equiv i\partial^{\mu}$ in 
the projectors must be replaced by their eigenvalues. 
This simple example shows what one can anticipate from
the application of the covariant mass-$m$ and spin-$\frac{3}{2}$ tracking
procedure to the four-vector spinor representation.

\section{Free spin-$\frac{3}{2}$ beyond the Rarita-Schwinger framework.}

In this Section we apply the spin-tracking procedure to the vector-spinor
representation. The decomposition of this space into Poincar\'e invariant
sectors reads%
\begin{equation}
\psi_{\mu}\longrightarrow
\lbrack 
\Psi^{\left(m,\frac{3}{2}\right) }\rbrack
^{(2)}\oplus\lbrack \Psi^{\left( m,\frac{1}{2}\right)}\rbrack ^{(4)},
\end{equation}
where the subscript labels the multiplicity of the representation.
Correspondingly, the Poincar\'e covariant spin-$\frac{3}{2}$ projector in
Eq.~(\ref{maxs_pr}) becomes
\begin{eqnarray}
\mathcal{P}^{(m;\frac{3}{2})}(p)&=&-\frac{1}{3}\left(  \frac{\mathcal{W}^{2}%
}{m^{2}}+\frac{3}{4}\frac{p^{2}}{m^{2}}\mathbf{1}_{16\times16}\right)  \,  
\end{eqnarray}
As long as the projectors are per construction covariant, 
the equation of motion
for spin-$\frac{3}{2}$ in $\psi_\mu$ and in any basis reads
\begin{equation}
\lbrack-\frac{1}{3}(\mathcal{W}^{2}+\frac{3}{4}p^{2}\mathbf{1}_{16\times16})-
m^{2}\mathbf{1}_{16\times16} ]\psi=0.
\label{bong}%
\end{equation}
In terms of the tensor $T_{AB\mu\nu}$, defined in Eq.~(\ref{W2}), the latter
equation rewrites to
\begin{equation}
\left[  -\frac{1}{3}T_{AB\mu\nu}p^{\mu}p^{\nu}-\left(  \frac{1}{4}p^{2}%
+m^{2}\right)  \delta_{AB}\right]  \psi^{B}=0,  \qquad
A:\alpha\,a,\quad B:\beta
\,b\,, \label{Platino}%
\end{equation}
where $a$ is the spinorial index.
%%%%%%%%%%%%%%%%%%%%%%%%%%%%%%%%
\subsection{The $\mathcal{W}^{2}$-- and $p^{2}$ driven spin-$\frac{3}{2}$
equations.}
In order to obtain the explicit form of Eq.~(\ref{Platino}) in the interesting
$\psi_{\mu}$ basis where Lorentz and spinor indices appear separated, we first
of all have to find $T_{AB\mu\nu}$, a calculation that we present in the
Appendix. Insertion of Eq.~(\ref{tensor}) from the Appendix into
Eq.~(\ref{Platino}) amounts to the following free spin-$\frac{3}{2}$ wave
equation
\begin{equation}
\lbrack -K_{\alpha\beta}+m^{2}~g_{\alpha\beta}%
]\psi^{\beta}=0\,, \label{concise}
\end{equation}
with
\begin{equation}
K_{\alpha\beta}\equiv\Gamma_{\alpha\beta\mu\nu}p^{\mu}p^{\nu},
\label{Platinoten}%
\end{equation}
where for the sake of simplicity we suppressed the spinorial indices and
defined $\Gamma_{\alpha\beta\mu\nu}$ as
\begin{equation}
\Gamma_{\alpha\beta\mu\nu} =\frac{2}{3}\left(  g_{\alpha\beta}g_{\mu\nu
}-g_{\alpha\nu}g_{\beta\mu}\right)
+\frac{1}{6}(\epsilon_{\quad\alpha\beta
\mu}^{\lambda}\gamma^{5}\sigma_{\lambda\nu}+\epsilon_{\quad\alpha\beta\nu
}^{\lambda}\gamma^{5}\sigma_{\lambda\mu})
+\frac{1}{12}\sigma_{\lambda\mu
}\sigma_{\quad\nu}^{\lambda}g_{\alpha \beta}-\frac{1}{4}g_{\mu\nu}
g_{\alpha\beta}\,.
\label{tensor_Gamma}
\end{equation}
It immediately verifies that the operator $K_{\alpha\beta}$ satisfies the
following relations%
\begin{equation}
p^{\alpha}K_{\alpha\beta}=0,\quad K_{\alpha\beta}p^{\beta}=0, \qquad
\gamma^{\alpha}K_{\alpha\beta}=0, \quad K_{\alpha\beta}\gamma^{\beta}=0.
\label{K_rel}%
\end{equation}
The resulting  free particle equation reads
\begin{equation}
{\Big[}  \left(  -p^{2}+m^{2}\right)  g_{\alpha\beta}+\frac{2}{3}p_{\beta
}p_{\alpha}
+\frac{1}{3}\left(  p_{\alpha}\gamma_{\beta}+p_{\beta}%
\gamma_{\alpha}\right)  \not p  -\frac{1}{3}\gamma_{\alpha}\not p
\gamma_{\beta}\not p  {\Big]}  \psi^{\beta}=0. \label{Platinoexplicit}%
\end{equation}
It equivalently rewrites as
\begin{equation}
{\mathcal P}^{\left(m,\frac{3}{2} \right) }_{\mu\nu}(p)\psi^\nu=
\frac{p^{2}}{m^{2}}{\mbox{\bf P}}_{\alpha\beta}^{(\frac{3}{2})}\,\psi^{\beta}%
=\psi_{\alpha}\,. \label{cov_prj_fct}%
\end{equation}
Here, ${\mbox{\bf P}}_{\alpha\beta}^{(\frac{3}{2})}$ 
stands  for the spin-$\frac{3}{2}$ projector 
in $\psi_{\mu}$ and is given by
\begin{eqnarray}
{\mbox{\bf P}}_{\alpha\beta}^{(\frac{3}{2})}&=&-\frac{1}{3}\left(  \frac
{\mathcal{W}_{\alpha\beta}^{2}}{p^{2}}+\frac{3}{4}g_{\alpha\beta}\right).
\label{3/2_projector}%
\end{eqnarray}
Equation (\ref{cov_prj_fct}) reveals the Poincar\'{e} invariant projector
\-
$\mathcal{P}^{\left( m, \frac{3}{2}\right)  }(p)$  as the
direct product of a mass-$m$--, and spin-$\frac{3}{2}$ projectors, much alike
Eq.~(\ref{revelation}) and as it should be. 
Using Eqs.~(\ref{K_rel}) allows to find  that the four-vector--spinor field
satisfies
\begin{eqnarray}
\lbrack p^{2}-m^{2}]\psi_{\alpha}  &  =&0,\label{NK}\\
\gamma_{\alpha}\psi^{\alpha}  &  =&0,\label{AC_1}\\
p_{\alpha}\psi^{\alpha}  &  =&0. \label{AC_2}%
\end{eqnarray}

\subsection{The spin-$\frac{3}{2}$ Lagrangian beyond Rarita-Schwinger.}
The equation of motion (\ref{concise}) can be derived from the following
manifestly Hermitian Lagrangian%
\begin{equation}
\mathcal{L}_{free}=-\frac{1}{2}[(\partial^{\mu}\overline{\psi}^{\alpha}%
)\Gamma_{\alpha\beta\mu\nu}\partial^{\nu}\psi^{\beta}+(\partial^{\nu}%
\overline{\psi}^{\beta})\overline{\Gamma}_{\alpha\beta\mu\nu}\partial^{\mu
}\psi^{\alpha}]
+m^{2}\overline{\psi}^{\alpha}\psi_{\alpha}\,, \label{KNlag}%
\end{equation}
where%
\begin{equation}
\overline{\Gamma}_{\alpha\beta\mu\nu}\equiv\gamma^{0}(\Gamma_{\alpha\beta
\mu\nu})^{\dagger}\gamma^{0}.
\end{equation}
Using Eq.~(\ref{tensor}) it is easy to show that
\begin{equation}
\overline{\Gamma}_{\alpha\beta\mu\nu}=\Gamma_{\beta\alpha\nu\mu}, 
\label{Eqchida}
\end{equation}
hence our Lagrangian can be rewritten to the simpler form%
\begin{equation}
\mathcal{L}_{free}=-(\partial^{\mu}\overline{\psi}^{\alpha})\Gamma_{\alpha
\beta\mu\nu}\partial^{\nu}\psi^{\beta}+m^{2}\overline{\psi}^{\alpha}%
\psi_{\alpha}. \label{NKlag}%
\end{equation}
Subjecting Eq.~(\ref{concise}) and its adjoint to standard algebraic
manipulations, or calculating directly the Noether current for the usual 
phase invariance of the Lagrangian (\ref{NKlag}) we obtain 
\begin{equation}
j_{\mu}=(i\partial^{\nu}\overline{\psi}^{\alpha})\Gamma_{\alpha\beta
\nu\mu}\psi^{\beta}-\overline{\psi}^{\alpha}\Gamma_{\alpha\beta\mu\nu}i\partial^{\nu}%
\psi^{\beta}\,, \label{Noether}%
\end{equation}
as a conserved current
\begin{equation}
\partial^{\mu}j_{\mu}=0.
\end{equation}
\subsection{The spin-$\frac{3}{2}$ propagator.}
The formal calculation of the two-point Green function in our theory requires
to work out the quantization of the formalism which is presently under
investigation and beyond the scope of this paper. However we calculated the
propagator as the inverse of the operator $\left( -K_{\alpha\beta
}+m^{2}g_{\alpha\beta}\right)$. In so doing, we obtain%

\begin{equation}
\Pi_{\alpha\beta}=\frac{\Delta_{\alpha\beta}}{\left(  p^{2}-m^{2}%
+i\varepsilon\right)  }\,,\label{3/2_PRJCT}%
\end{equation}
where the projector over "positive energy" solutions turns out to be
\begin{equation}
\Delta_{\alpha\beta}=-g_{\alpha\beta}+\frac{2}{3m^{2}}p_{\beta}p_{\alpha
}+\frac{1}{3m^{2}}\left(  p_{\alpha}\gamma_{\beta}+p_{\beta}\gamma_{\alpha
}\right)  \not p
-\frac{1}{3m^{2}}\gamma_{\alpha}\not p\gamma_{\beta}\not p\,.
\end{equation}
It is instructive to rewrite this tensor in terms of the projectors over well
defined spins. The result is
\begin{equation}
\Delta_{\alpha\beta}=-{\mbox{\bf P}}_{\alpha\beta}^{(\frac{3}{2})}+\frac
{p^{2}-m^{2}}{m^{2}}{\mbox{\bf P}}_{\alpha\beta}^{(\frac{1}{2})}%
\,,\label{3/2_prop}%
\end{equation}
with
\begin{equation}
{\mbox{\bf P}}_{\alpha\beta}^{(\frac{1}{2})}=
\frac{\mathcal{W}_{\alpha\beta}^{2}%
}{3p^{2}}+\frac{5}{4}g_{\alpha\beta},
\label{prj12}
\end{equation}
being the projector on spin-$\frac{1}{2}$ in $\psi_{\mu}$.
Equation (\ref{3/2_prop}) shows that off-shell the four-vector spinor carries
all its lower spin components, much alike the case of the four-vector 
in the description of  ``off-shell''  electroweak gauge bosons 
\cite{Mohamed}.

That $\Pi_{\alpha\beta}$ is the inverse to the free particle equation can be 
easily shown using Eq.~(\ref{cov_prj_fct}) in combination with the nilpotent 
and orthogonality properties of the projectors. 
The similarity of the spin-$\frac{3}{2}$ propagator in Eq.~(\ref{3/2_prop}) 
and the $A_{\mu}$-propagator in Eq.~(\ref{prop_1}) can hardly be overlooked. 

\noindent Finally, the $\Delta_{\alpha\beta}$ operators have the following
simple properties:%
\begin{eqnarray}
p^{\alpha}\Delta_{\alpha\beta}=\frac{1}{m^{2}}\left(  p^{2}-m^{2}\right)
p_{\beta},  &  \quad& \Delta_{\alpha\beta}p^{\beta}=\frac{1}{m^{2}}\left(
p^{2}-m^{2}\right)  p_{\alpha}\,,\nonumber\\
\gamma^{\alpha}\Delta_{\alpha\beta}=\frac{1}{m^{2}}\left(  p^{2}-m^{2}\right)
\gamma_{\beta},  &  \quad& \Delta_{\alpha\beta}\gamma^{\beta}=\frac{1}{m^{2}%
}\left(  p^{2}-m^{2}\right)  \gamma_{\alpha}. \label{Delta_prpt}\nonumber\\
\end{eqnarray}

Before concluding the current section, we wish to notice that
Eq.~(\ref{cov_prj_fct}) finds a marginal mentioning in Ref.~\cite{nieu}
however without any discussion on its link to $\mathcal{W}^2$ 
and without exploiting its potential in the description of
spin-$\frac{3}{2}$.

%%%%%%%%%%%%%%%%%%%%%%%%%%%%%%%%%%%%%%%%%%

\section{Symmetries of the Lagrangian.}
This section is devoted to the symmetries of
Eq.~(\ref{concise}) in the massless limit, their impact
on the massive case and its relation to the prime question of the uniqueness 
of the Lagrangian in Eq.~(\ref{NKlag}).

\subsection{Parameter independence of the massless case.}
In the massless case Eq.~(\ref{concise}) remains invariant 
under the following ``gauge'' transformation:
\begin{equation}
\psi^{\beta}\rightarrow\psi^{\prime\beta}=\psi^{\beta}+p^{\beta}\chi,
\label{Gl_1}
\end{equation}
with $\chi$ being an arbitrary spinor. This invariance 
appears as a consequence of
\begin{equation}
 K_{\alpha\beta}p^{\beta}=0. \label{prop1}%
\end{equation}
It is same as the conventional ``gauge'' symmetry satisfied also by 
the ($A=-1$) version of the massless RS equation of motion in 
Eq.~(\ref{super}) which has been extensively used in particular in 
supergravity \cite{nieu}. More recently,
the  symmetry in Eq.~(\ref{Gl_1}) 
has been exploited  as a guiding principle in the construction of 
chiral Lagrangians for light baryons \cite{Pascaft}. 
The relation 
\begin{equation}
K_{\alpha\beta}\gamma^{\beta}=0, \label{prop2}%
\end{equation}
in Eq.~(\ref{K_rel}) implies  invariance of  
Eq.~(\ref{concise}) under the point transformation 
\begin{equation}
\psi^{\beta}\rightarrow\psi^{\prime\beta}=\psi^{\beta}+\gamma^{\beta}\chi .
\label{point2}
\end{equation}
At that stage, the question on the uniqueness of the 
formalism proposed here comes up. 
In order to answer this question, let us first
perform the following "rotation" within the 
unphysical spin-$\frac{1}{2}$ sector 
\begin{equation}
\psi^{\beta}\rightarrow\psi^{\prime\beta}=
R^{\beta\rho}\left(\frac{A}{2}\right)\psi^{\rho},
\end{equation}
with $R(w)$ given by  Eq.~(\ref{rotation}).
In so doing one produces the $A$-dependent Lagrangian
\begin{equation}
\mathcal{L}(A)   =-(\partial^{\mu}\overline{\psi^\prime }\acute{}
^{\alpha})\Gamma_{\alpha\beta\mu\nu}\partial^{\nu}\psi^{\prime\beta}
= -\left(  \partial^{\mu}\overline{\psi}^{\sigma}\right)  \Gamma_{\sigma
\rho\mu\nu}(A)\partial^{\nu}\psi^{\rho},
\end{equation}
with
\begin{equation}
\Gamma_{\sigma\rho\mu\nu}(A)  =
R_{\sigma}^{\quad\alpha}\left( \frac{A}{2}\right)\Gamma
_{\alpha\beta\mu\nu}R_{\quad\rho}^{\beta}\left(\frac{A}{2}\right)\,.
\end{equation}
At first glance, the equation of motion resulting from the latter
 Lagrangian presents itself $A$-dependent as
\begin{equation}
 -\Gamma_{\sigma\rho\mu\nu}(A)p^{\mu}p^{\nu}  \psi^{\rho}=0.
\end{equation}
However this impression is misleading. Indeed, in making use
of Eq.~(\ref{prop2}) allows to eliminate the $A$-dependence according to
\begin{equation}
K_{\sigma\rho}(A)\equiv\Gamma_{\sigma\rho\mu\nu}(A)p^{\mu}p^{\nu}
=R_{\sigma}^{\quad\alpha}\left(\frac{A}{2}\right)K_{\alpha\beta}
R_{\quad\rho}^{\beta}\left(\frac{A}{2}\right)=K_{\sigma\rho}.
\label{a_free}
\end{equation}
This means that in the massless limit our Eq.~(\ref{concise}) is  unique. 
This uniqueness is of course related to the invariance under the point 
transformations in Eq.~(\ref{point2}) as can be easily seen in choosing
the $\chi $ spinor as
$\chi =\frac{A}{2}\gamma\cdot \psi $.
The only way for two different Lagrangians to produce one and the 
same equation of motion is that they differ  by a total divergence term
according to
\begin{equation}
\mathcal{L}(A)=\mathcal{L}-\partial^{\mu}\Lambda_{\mu}(A).
\end{equation}
That this is indeed the case follows directly from the explicit 
calculation of $\Lambda_{\mu}(A)$ giving 
\begin{equation}
\Lambda_{\mu}(A)=\frac{A}{2}\left[  \overline{\psi}^{\sigma}\gamma_{\sigma}%
\gamma^{\alpha}\Gamma_{\alpha\rho\mu\nu}\partial^{\nu}\psi^{\rho}+\left(
\partial^{\tau}\overline{\psi}^{\sigma}\right)  \Gamma_{\sigma\beta\tau\mu
}\gamma^{\beta}\gamma_{\rho}\psi^{\rho}\right]  
+\frac{A^{2}}{4}\overline{\psi}^{\sigma
}\gamma_{\sigma}\gamma^{\alpha}\Gamma_{\alpha\beta\mu\nu}\gamma^{\beta}%
\gamma_{\rho}\partial^{\nu}\psi^{\rho}.
\end{equation}
In this manner the parameter independence of our suggested formalism in 
the massless case establishes neatly.
Notice however that the operator $K_{\alpha\beta}$ is 
not invertible, meaning that the propagator in Eq.~(\ref{3/2_PRJCT}) is 
singular in the massless case. Same occurs for spin-$1$ due
to the singularity of the massless operator in Eq.~(\ref{proca_pak}). 
This problem reflects the 
gauge freedom of massless theories and is resolved by  introducing a gauge 
fixing term into the Lagrangian, a technique that will acquire importance
in the following.

\subsection{Extrapolation to the massive case.} 
The mass term in Eq.~(\ref{concise}) breaks both the
gauge symmetry and the invariance under point transformations 
and is defined
in the respective Eqs.~(\ref{prop2},\ref{point2}). 
In the massive case calculations similar to those presented 
in the previous subsection, yield the following genuinely
$A$-dependent Lagrangian
\begin{equation}
\mathcal{L}(A)=\mathcal{L}_{K}+\overline{\psi}^{\sigma}\left[
M^{2}(A)\right]_{\sigma\rho}\psi^{\rho}-\partial^{\mu}\Lambda_{\mu}(A),
\end{equation}
with
\begin{equation}
\left[  M^{2}(A)\right]  _{\sigma\rho}=m^{2}
R_{\sigma}^{\quad\alpha}\left(\frac{A}{2}\right)
R_{\alpha\rho}\left( \frac{A}{2}\right )=m^{2}R_{\sigma\rho}(A(1+A)).
\end{equation}
In order to understand the effect of the "rotation" within the 
spin-$\frac{1}{2}$ sector, let us rewrite the equation of motion in terms 
of the projectors as
\begin{equation}
\lbrack -p^{2}({\mbox{\bf P}}^{\left(\frac{3}{2}\right)})_{\mu\nu}+
m^{2}g_{\mu\nu}]\psi^{\nu
}=0.\label{projector_Gl}
\end{equation}
Next we shall separate the spin-$\frac{3}{2}$-- from the spin-$ \frac{1}{2}$ 
mass term as
\begin{equation}
\lbrack\left(  -p^{2}+m^{2}\right)  ({\mbox{\bf P}}
^{\left( \frac{3}{2}\right)})_{\mu\nu
}+m^{2}({\mbox{\bf P}}^{\left(\frac{1}{2}\right)})_{\mu\nu}]\psi^{\nu}=0.
\end{equation}
The $A$-dependent equation of motion  can be written as%
\begin{equation}
R\left( \frac{A}{2}\right) [\left(  -p^{2}+m^{2}\right)  
{\mbox{\bf P}}^{\left(\frac{3}{2}\right)}+m^{2}{\mbox{\bf P}}%
^{\left(\frac{1}{2}\right)}]R\left( \frac{A}{2}\right)\psi=0\, . 
\label{rotatedeom}
\end{equation}
It is convenient now to use the ${\mbox{\bf P}}^{\left(\frac{3}{2}\right)}$, 
${\mbox{\bf P}}_{11}^{\left(\frac{1}{2}\right)}$, and 
${\mbox{\bf P}}_{22}^{\left(\frac{1}{2}\right)}$  
projectors, and the so called "switch" operators,
${\mbox{\bf P}}_{12}^{\left(
\frac{1}{2}\right)}, 
{\mbox{\bf P}}_{12}^{\left( 
\frac{1}{2}
\right)}$ which can be found, among others, in Ref. \cite{nieu} and read
\begin{eqnarray}
({\mbox{\bf P}}^{\left( \frac{3}{2}\right)})_{\mu\nu}  & =&
g_{\mu\nu}-\frac{1}{3}\gamma_{\mu}\gamma
_{\nu}-\frac{1}{3p^{2}}(\not p \gamma_{\mu}p_{\nu}+p_{\mu}\gamma_{\nu}%
\not p ),\nonumber \\
({\mbox{\bf P}}_{11}^{\left(\frac{1}{2}\right)})_{\mu\nu}  &=&-
\frac{p_{\mu}p_{\nu}}{p^{2}}+\frac
{1}{3}\gamma_{\mu}\gamma_{\nu}+\frac{1}{3p^{2}}(\not p \gamma_{\mu}p_{\nu
}+p_{\mu}\gamma_{\nu}\not p ),\nonumber\\
\left(
{\mbox{\bf P}}_{22}^{
\left( 
\frac{1}{2}
\right)}
\right)_{\mu\nu}  & =&
\frac{p_{\mu}p_{\nu}}{p^{2}},\nonumber\\
\left({\mbox{\bf P}}_{12}^{\left( \frac{1}{2}\right)}\right)_{\mu\nu}  
&  =&\frac{1}{\sqrt{3}p^{2}}
\left(p_{\mu}p_{\nu
}-\not p \gamma_{\mu}p_{\nu}\right),\nonumber\\
\left({\mbox{\bf P}}_{21}^{\left( \frac{1}{2}\right)}\right)_{\mu\nu}  &  =&
\frac{1}{\sqrt{3}p^{2}}\left(-p_{\mu}p_{\nu
}+\not p p_{\mu}\gamma_{\nu}\right). \label{VAR_PROJ}
\end{eqnarray}
Above operators constitute a complete set in the vector-spinor representation 
space and satisfy the following orthogonality and completeness relations%
\begin{equation}
{\mbox{\bf P}}_{ij}^{a}{\mbox{\bf P}}_{kl}^{b}    =\delta^{ab}\delta_{jk}%
{\mbox{\bf P}}_{il}^{b}, \qquad
{\mbox{\bf P}}^{\left(\frac{ 3}{2}\right)}+
{\mbox{\bf P}}_{11}^{\left( \frac{1}{2}\right)}+
{\mbox{\bf P}}_{22}^{\left( \frac{1}{2}\right)}  =1. \label{pp}
\end{equation}
Further  useful relations are%
\begin{eqnarray}
\not p {\mbox{\bf P}}^{\left( \frac{3}{2}\right)}  &  =&
{\mbox{\bf P}}^{\left( \frac{3}{2}\right)}\not p, \qquad
\not p {\mbox{\bf P}}_{ij}^{\left(\frac{1}{2}\right)}    =
\pm{\mbox{\bf P}}_{ij}^{\left(\frac{ 1}{2}\right)}\not p, \qquad%
\begin{tabular}
[c]{l}%
$+$ if $i=j$\\
$-$ if $i\neq j$%
\end{tabular},
\nonumber\\
\gamma^{\mu}{\mbox{\bf P}}_{\mu\nu}^{\left( \frac{3}{2}\right)}  &  =&
{\mbox{\bf P}}_{\mu\nu}^{\left(\frac{ 3}{2}\right)}%
\gamma^{\nu}=p^{\mu}{\mbox{\bf P}}_{\mu\nu}^{\left(\frac{3}{2}\right)}=
{\mbox{\bf P}}_{\mu\nu}^{\left(\frac{ 3}{2}\right)}p^{\nu
}=0,\nonumber\\
\gamma^{\mu}\left({\mbox{\bf P }}
^{\left( \frac{1}{2}\right)}\right)_{\mu\nu}  &  =&
\gamma_{\nu}, \qquad \left({\mbox{\bf P}}%
^{\left( \frac{1}{2}\right)}\right)_{\mu\nu}\gamma^{\nu}=
\gamma_{\mu}, \qquad
p^{\mu}\left({\mbox{\bf P}}^{\left(\frac{1}{2}\right)}\right)_{\mu\nu}    =
p_{\nu}, \qquad \left({\mbox{\bf P}}^{\left( \frac{1}{2}\right)}%
\right)_{\mu\nu}p^{\nu}=p_{\mu}.\label{rproj}
\end{eqnarray}
The 
${\mbox{\bf P}}^{\left( \frac{3}{2}\right) }$ projector
was related to the squared Pauli-Lubanski
vector in Eq.~(\ref{3/2_projector}) 
whereas the projector in Eq.~(\ref{prj12}) expresses as
\begin{equation}
{\mbox{\bf P}}^{\left( \frac{1}{2}\right)}=
{\mbox{\bf P}}_{11}^{\left( \frac{1}{2}\right)}+{\mbox{\bf P}}_{22}
^{\left( \frac{1}{2} \right) }.
\label{prj12:2}
\end{equation}
These relations can be exploited  to cast Eq.~(\ref{rotatedeom}) into
the form which manifestly shows that
{\it solely the mass term  in the spin-$\frac{1}{2}$ sector
is affected  by the point transformation\/},
\begin{equation}
\lbrack\left(  -p^{2}+m^{2}\right)  
{\mbox{\bf P}}^{\left(\frac{3}{2}\right)}+m^{2}
R\left( \frac{A}{2}\right){\mbox{\bf P}}%
^{\left( \frac{1}{2}\right)}
R\left( \frac{A}{2}\right)]\psi=0. \label{eomA}
\end{equation}
Under the above ``rotation'' the mass matrix changes  
from a diagonal form in both spin-sectors
to one that remains diagonal only in the spin-$\frac{3}{2}$--
but becomes  non-diagonal in  
the spin-$\frac{ 1}{2}$ sector. 
This non-diagonality is irrelevant for on-shell 
particles as is well visible  upon
contracting Eq.~(\ref{eomA}) with $\gamma^{\sigma}$ and $p^{\sigma}$, 
and recalling that the invertibility of $R\left(\frac{A}{2}\right)$ 
requires $A\neq-\frac{1}{2}$,
\begin{eqnarray}
m^2(1+2A)^2\gamma\cdot\psi=0\, ,
&\Rightarrow& \gamma\cdot\psi=0\,, \nonumber \\
m^2(p\cdot\psi+A(A+1)\gamma\cdot\psi)=0,
&\Rightarrow& p\cdot\psi=0.
\end{eqnarray} 
In this fashion, the form of the mass matrix in 
the spin-$\frac{1}{2}$ sector in the free equation
remains without importance. However, it becomes 
relevant for the  off-mass shell propagator.
In order to see this, notice, that the
$A-$dependent propagator is easily calculated in the following way.
Let us first denote $(-K_{\alpha\beta}+m^2g_{\alpha\beta})$ in 
Eq.~(\ref{concise}) by ${\mathcal O}_{\alpha \beta}$.
Upon the $R\left(\frac{A}{2}\right)$ transformation, 
${\mathcal O}_{\alpha \beta}$
becomes ${\mathcal O}(A)=R\left( \frac{A}{2}\right){\mathcal O}
R\left( \frac{A}{2}\right)$ 
and as long as the new propagator comes from
$\Pi (A) {\mathcal O}(A)=1$ then 
one finds  $\Pi (A)$ from 
\begin{equation}
 R\left(\frac{A}{2}\right)^{-1}\Pi 
R\left( \frac{A}{2}\right)^{-1}
R\left( \frac{A}{2}\right)
{\mathcal O}R\left(\frac{A}{2}\right)=1\, .
\label{Gl_2}
\end{equation}
The latter equation leads to the following $A$-dependent massive
propagator, 
\begin{equation}
\Pi\left(A\right)=\frac{-{\mbox{\bf P}}^{\left(\frac{3}{2}\right)}+
\frac{p^{2}-m^{2}}{m^{2}}R^{-1}%
\left(\frac{A}{2}\right){\mbox{\bf P}}^{\left(\frac{1}{2}\right)}
R^{-1}\left(\frac{A}{2}\right)}{p^{2}-m^{2}+i\epsilon }.
\label{propA}%
\end{equation}
Therefore, the  remnant $A$-dependence affects 
only the spin-$\frac{ 1}{2 }$ contribution to the 
off-shell propagator. In other words, the parameter dependence
of the massive spin-$\frac{3}{2}$ propagator appeared as a consequence
of respecting in the massive theory the symmetries of the massless
case. The situation is by no means new.
In a similar way, the gauge symmetry of the massless spin-$1$ 
theory is respected by the massive one on the cost of a 
parameter dependent spin-$0$ sector in  the massive gauge boson
propagator, a subject that we treat in some detail  the following
subsection.

\subsection{Off-shell propagators and parameter dependence.}

\noindent
In this subsection we shall make the case that 
the massive off-shell spin-$\frac{3}{2}$ propagator proposed
here is of the type of the propagators which appear in massive gauge theories
and that its parameter dependence reflects the symmetries of the massless
theory, one of them being the gauge freedom.
In order to see this we begin with casting the essentials of the 
standard massive gauge theories in the language
systematically used by us through this paper, namely the one 
of the covariant projectors as applied to the spin-$0$--, and spin-$1$
sectors in the $(\frac{1}{2},\frac{1}{2})$ space. Then we analyze     
the parameter dependence of our propagator in the light of the symmetries 
of the massless Lagrangian. 
 
\subsubsection{ Gauge fixing in the massive spin-$1$ propagator.}

The problem of the parameter dependence of the off-shell propagators
is quite general indeed and appears in massive spin-$1$ gauge theories. 
In the massless case, the equation ~(\ref{proca_pak}) is not invertible 
as visible from  Eqs.~(\ref{Proca-Prop}), (\ref{prop_1}).
The non-invertibility reflects the gauge freedom and it is circumvented 
by the introduction of appropriate gauge fixing terms into the 
Lagrangian according to:
\begin{equation}
\mathcal{L}_{QED}=-\frac{1}{4}F^{\mu\nu}F_{\mu\nu}
-J^{\mu}A_{\mu} -\frac{1}{2a}\left(\partial^{\mu}A_{\mu}\right)  ^{2}.
\label{Gl_3}
\end{equation}
Here, the $J^{\mu}$ current depends on the matter fields. 
The wave equation associated with
the latter Lagrangian reads
\begin{equation}
\left[- p^{2}g_{\mu\nu}-\left(  \frac{1}{a}-1\right)  p_{\mu}p_{\nu}\right]
A^{\nu}=J_{\mu}\,.\label{QEDa}%
\end{equation}
Equation (\ref{QEDa}) is now invertible and leads to the following
$a$-dependent  propagator%
\begin{equation}
\Pi_{\mu\nu}(a)=\frac{1}{p^{2}+i\varepsilon}\left[ - g_{\mu\nu}%
+(1-a)\frac{p^{\mu}p^{\nu}}{p^{2}}\right]  
=\frac{1}{p^{2}+i\varepsilon
}\left[  -{\mbox{\bf P}}_{\mu\nu}^{(1)}-
a{\mbox{\bf P}}_{\mu\nu}^{(0)}\right],\label{propa}%
\end{equation}
with the spin-$1$ projectors from Eq.~(\ref{proj1}). 
Choosing specific values for $a$ is standard and known as
``gauge fixing''. The  $a=1$ value in Eq.~(\ref{propa}) is known as
Feynman's gauge while $a=0$ gives the Landau propagator. 

The link to the massive case is established by noticing that
the gauge condition becomes a constraint that is preserved under 
interactions whenever the massive gauge boson is coupled to a 
conserved current. Such is possible only within the context
of mass-generation via the Higss mechanism, a possibility 
which we highlight in brief in what follows. 
To be specific, in massive gauge theories one faces the problem to
guarantee validity of the gauge condition $\partial \cdot A=0$.
{}For this purpose and in analogy to the massless theory
one introduces  a Lagrange multiplier into the Lagrangian according to 
\begin{equation}
\mathcal{L}_{P}=-\frac{1}{4}F^{\mu\nu}F_{\mu\nu}+\frac{m^{2}}{2}A^{\mu}A_{\mu}-\frac
{1}{2a}\left(  \partial^{\mu}A_{\mu}\right)  ^{2}-J^{\mu}A_{\mu}.
\end{equation}
The resulting massive equation of motion now becomes%
\begin{equation}
\left[  \left(  -p^{2}+m^{2}\right)  g_{\mu\nu}-\left(  \frac{1}{a}-1\right)
p_{\mu}p_{\nu}\right]  A^{\nu}=J_{\mu}\, ,%
\end{equation}
or equivalently,%
\begin{equation}
\left[  \left(  -p^{2}+m^{2}\right)  {\mbox{\bf P}}_{\mu\nu}^{(1)}-\frac{1}%
{a}\left(  p^{2}-am^{2}\right)  {\mbox{\bf P}}_{\mu\nu}^{(0)}\right]  A^{\nu
}=J_{\mu}\, .%
\end{equation}
Here, $p_{\mu}$ denotes the four momentum of the gauge boson.
The associated  propagator is well known and obtained as 
\begin{equation}
\Pi_{\mu\nu} (a)=\frac{1}{p^{2}-m^{2}+i\varepsilon}\left[
{-\mbox{\bf P}}_{\mu\nu}^{(1)}-a\frac{p^{2}-m^{2}}{p^{2}-a
m^{2}}{\mbox{\bf P}}_{\mu\nu
}^{(0)}\right] 
=\frac{-g_{\mu\nu}+(1-a)\frac{p^{\mu}p^{\nu}}{p^{2}-am^{2}}%
}{p^{2}-m^{2}+i\varepsilon}.\label{proppa}%
\end{equation}
The Proca propagator in Eq.~(\ref{Proca-Prop})
corresponds to the particular choice of  $a=\infty$,
and appears  singular in the massless case, much
alike our propagator in Eq.~(\ref{3/2_PRJCT}). 

Although not obvious, the latter expression is related to 
the conventional mass-generation mechanism for gauge bosons
via the Higgs mechanism \cite{AH}. In order to 
illustrate this statement one couples the gauge boson
to a charged scalar ("Higgs") field defined as 
\begin{equation}
\phi=\frac{1}{\sqrt{2}}(v+\chi_{1}+i\chi_{2})\, ,
\label{Higgs_bos}
\end{equation}
and obtains the equation of motion from%
\begin{eqnarray}
\partial^{\mu}F_{\mu\nu} &  =&j_{\nu}=e\left[  \phi^{\ast}(i\partial_{\nu}\phi)
-(i\partial_{\nu} \phi^{\ast})\phi\right]  -2e^{2}A_{\nu}
\phi^{\ast}\phi \nonumber\\
&  =&-m^{2}A_{\nu}-m \partial_{\nu}\chi_{2}-
e\left[  \chi_{1}\partial_{\nu}\chi_{2}
-\chi_{2}\partial_{\nu}\chi_{1}\right] 
 - e^{2}A_{\nu}(\chi_{1}^{2}+2v\chi_{1}+\chi_{2}%
^{2})\, ,\label{eomg}
\end{eqnarray}
with $m\equiv ev$. In considering  now the special gauge
\begin{equation}
\partial_{\mu}A^{\mu}=m\xi\chi_{2}\, , \label{chi}
\end{equation}
where $\xi$ is an arbitrary parameter one is led to
\begin{equation}
\chi_{2}=\frac{1}{m\xi}\partial_{\mu}A^{\mu}\, .%
\end{equation}
With that the equation of motion for the gauge boson becomes%
\begin{equation}
\left[ -p^{2}+m^{2}\right]  A_{\nu}-(\frac{1}{\xi}-1)p_{\nu}p^{\mu}A_{\mu}
= ie\left[  \chi_{1}p_{\nu}\chi_{2}-\chi_{2}p_{\nu}\chi_{1}\right] 
 -e^{2}A_{\nu}(\chi_{1}^{2}+2v\chi_{1}+\chi_{2}^{2}).
\end{equation}
The right hand side of the latter equation contains interactions of the 
gauge boson  field with the
Higgs field, $\chi_{1}$, as well as self-interactions. 
Its left hand side can be inverted to
yield the well known 't Hooft propagator
\begin{equation}
\Pi_{\mu\nu}^{'t Hooft}(\xi)=\frac{-g_{\mu\nu}+(1-\xi)\frac{p^{\mu}p^{\nu}
}{p^{2}-\xi m^{2}}}{p^{2}-m^{2}+i\varepsilon} \label{tHooft_prop} \\
=\frac{1}{p^{2}-m^{2}
+i\varepsilon}\left[ - {\mbox{\bf P}}_{\mu\nu}^{(1)}-\xi\frac{p^{2}-m^{2}}
{p^{2}-\xi m^{2}}{\mbox{\bf P}}_{\mu\nu}^{(0)}\right].
\nonumber
\end{equation}
The 't Hooft  propagator describes massive vector particles
whose  mass has been generated via
the Higgs mechanism. Now one recovers the propagator in Eq.~(\ref{proppa})
in assuming $v\neq 0$, and $\xi=a$ in Eqs.~(\ref{Higgs_bos},\ref{chi}).
The  $v=0$ value  implies  $m=0$ 
and the absence of spontaneous 
symmetry breaking. The resulting massless propagator coincides with the one 
given in Eq.~(\ref{propa}). 
This brief  reminiscence of the massive spin-$ 1$ case 
is suggestive of the idea to view the $A$-dependence of 
the massive spin-$\frac{3}{2}$ propagator in Eq.~(\ref{propA}) 
in the light of gauge fixing, an idea that we execute in the next subsection.

\subsubsection{"Gauge" fixing in the spin-$\frac{3}{2}$ theory.}
In parallel to the spin-$1$ description, we here shall
term to the freedom in the choice of the massless $\psi^{\beta}$
provided by the symmetries in Eqs.~(\ref{Gl_1}), and (\ref{point2}) 
as ``gauge'' freedom (better, ``gauge'' freedoms), and
fix it by including into the Lagrangian the
associated Lagrange multipliers according to
\begin{equation}
\mathcal{L}=-(\partial^{\mu}\overline{\psi}^{\alpha})
\Gamma_{\alpha\beta\mu\nu
}\partial^{\nu}\psi^{\beta}\,
-\frac{1}{a}\left(  \partial^{\mu}\overline{\psi}_{\mu}\right)  \left(
\partial^{\alpha}\psi_{\alpha}\right)  
-\frac{\mu^{2}}{b}\left(  \overline{\psi
}_{\mu}\gamma^{\mu}\right)  \left(  \gamma^{\alpha}\psi_{\alpha}\right)
-\overline{\psi}^{\mu}f_{\mu}-\overline{f}^{\mu}\psi_{\mu}.
\label{Lagra}
\end{equation}
Notice that we use the Lagrange multiplier 
$\mu^{2}/b$ where $\mu$ is an arbitrary 
(but fixed) mass scale which allows to treat the parameters
as dimensionless. This new  Lagrangian 
yields now the equation of motion as%
\begin{equation}
\left[  -K_{\alpha\beta}\,-\frac{1}{a}p_{\alpha}p_{\beta
}-\frac{\mu^{2}}{b}\gamma_{\alpha}\gamma_{\beta}\right]  \psi^{\beta}=
f_{\alpha
}\, ,\label{eomabm0}%
\end{equation}
where $f_\mu$ is some fermion current involving other fields. The operator on 
the left hand side of the latter equation is now invertible.
In terms of the projectors in 
Eq.~(\ref{VAR_PROJ}) it is given by
\begin{equation}
{\mathcal O}^{(m=0)}(a,b)= - p^{2}{\mbox{\bf P}}^{\left(\frac{3}{2}\right)} 
-\frac{3\mu^{2}}{b}{\mbox{\bf P}}_{11}^{\left(\frac{1}{2}\right)} 
-\frac{1}{ab}\big( b p^{2}+a\mu^{2}\big) {\mbox{\bf P}}_{22}^{1/2}
 -\frac{\sqrt{3}\mu^{2}}{b}({\mbox{\bf P}}_{12}^{\left( \frac{1}{2}\right)}+
{\mbox{\bf P}}_{21}^{\left( \frac{1}{2} \right) }),
\end{equation}
where we used
\begin{equation}
\gamma_{\alpha}\gamma_{\beta}=\left[ 3{\mbox{\bf P}}_{11}^
{\left(\frac{1}{2}\right)}+{\mbox{\bf P}}_{22}%
^{\left( \frac{1}{2}\right)}+\sqrt{3}({\mbox{\bf P}}_{12}
^{\left( \frac{1}{2}\right) }+
{\mbox{\bf P}}_{21}^{\left(\frac{1}{2}\right)})\right]_{\alpha\beta}.
\end{equation}
The propagator in the "$(a,b)$-gauge" is now found to be  
\begin{equation}
\Pi^{(m=0)}(a,b)=\frac{\Delta^{(m=0)}(a,b)}{p^{2}+i\varepsilon}\, .
\label{propabm0}
\end{equation}
Here,
\begin{equation}
\Delta^{(m=0)}(a,b)=-{\mbox{\bf P}}^{\left(\frac{3}{2}\right)}-
\frac{1}{3}{\big[}\left(  \frac{b}{\mu^2} p^{2}+a\right)
{\mbox{\bf P}}_{11}^{\left(\frac{1}{2}\right)} 
 +3a{\mbox{\bf P}}_{22}^{\left(\frac{1}{2}\right)}-
\sqrt{3}a({\mbox{\bf P}}%
_{12}^{\left( \frac{1}{2}\right) }+{\mbox{\bf P}}_{21}
^{\left( \frac{1}{2} \right)}){\big]}.
\end{equation}
This is the spin-$\frac{3}{2}$ analogous to the massless spin-$1$  
propagator in Eq.~(\ref{propa}).

Next we extrapolate to the massive case. Adding the mass term to the 
Lagrangian results in
\begin{equation}
\mathcal{L}=-(\partial^{\mu}\overline{\psi}^{\alpha})
\Gamma_{\alpha\beta\mu\nu
}\partial^{\nu}\psi^{\beta}\,
+m^{2}\overline{\psi}^{\alpha}\psi_{\alpha}-
\frac{1}{a}\left(  \partial^{\mu}\overline{\psi}_{\mu}\right)  \left(
\partial^{\alpha}\psi_{\alpha}\right) 
 -\frac{\mu^{2}}{b}\left(  \overline{\psi
}_{\mu}\gamma^{\mu}\right)  \left(  \gamma^{\alpha}\psi_{\alpha}\right)
-\overline{\psi}^{\mu}f_{\mu}-\overline{f}^{\mu}\psi_{\mu}.
\label{Lagra_mass}
\end{equation} 
The massive equation of motion reads
\begin{equation}
\left[  -K_{\alpha\beta}+m^{2} g_{\alpha\beta}\,-\frac{1}{a}p_{\alpha}p_{\beta
}-\frac{\mu^{2}}{b}\gamma_{\alpha}\gamma_{\beta}\right]  \psi^{\beta}=
f_{\alpha
}\, .\label{eomab}%
\end{equation}
In the  $a=\infty$-''gauge''  
Eq.~(\ref{eomab}) corresponds to the "rotated"
Eq.~(\ref{eomA}), modulo the identification 
$\frac{\mu^2}{b}=-A(1+A)m^{2}$. 
{\it This observation reveals the effect of the rotation in Eq.~(\ref{eomA}) 
just as a change in the "gauge" used in the massless case 
(massive case, in reference to the Higgs mechanism)}. 

In terms of the projectors in Eq.~(\ref{VAR_PROJ}), the operator acting on 
the field on the left hand side in Eq.~(\ref{eomab})  reads 
\begin{equation}
{\mathcal O}(a,b)=\left( - p^{2}+m^{2}\right)  
{\mbox{\bf P}}^{\left(\frac{3}{2}\right)}-\frac{1}{b}%
(3\mu^{2}-bm^{2}){\mbox{\bf P}}_{11}^{\left(\frac{1}{2}\right)}
-\frac{1}{ab}\big( b p^{2}+a(\mu^{2}-bm^{2})\big) {\mbox{\bf P}}_{22}^{1/2}
 -\frac{\sqrt{3}\mu^{2}}{b}({\mbox{\bf P}}_{12}^{\left( \frac{1}{2}\right)}+
{\mbox{\bf P}}_{21}^{\left( \frac{1}{2} \right) }).
\end{equation}
This operator has an inverse which is calculated  as%
\begin{equation}
\Pi(a,b)=\frac{\Delta(a,b)}{p^{2} -m^{2}+i\varepsilon}\label{propab}\, ,
\end{equation}
with
\begin{eqnarray}
\Delta(a,b)&=&
-{\mbox{\bf P}}^{\left(\frac{3}{2}\right)}-b\frac{p^{2}-m^{2}}
{(3\mu^{2}-bm^{2})(b p^{2}+a(\mu^{2}-bm^{2}))-3a\mu^{4}}\times \nonumber \\
&& \left[  \left(  b p^{2}+a(\mu^{2}-bm^{2})\right){\mbox{\bf P}}_{11}
^{\left(\frac{1}{2}\right)}
+a\left(  3\mu^{2}-b m^{2}\right)  {\mbox{\bf P}}_{22}
^{\left(\frac{1}{2}\right)} 
-\sqrt{3}a\mu^{2} ({\mbox{\bf P}}_{12}^{(\frac{1}{2})}+
{\mbox{\bf P}}_{21}^{(\frac{1}{2}) }) \right] \, .  
\label{Delab} %
\end{eqnarray}
The similarity of the massive spin-$\frac{3}{2}$ off-shell propagator
in Eq.~(\ref{propab}) with the nominator from 
Eq.~(\ref{Delab}) to the t'Hooft propagator in 
Eq.~(\ref{tHooft_prop}) is hardly to be overlooked.
In both cases the parameter dependence invokes propagation
of the unphysical spin-sectors that have been excluded on-shell. 
This observation is suggestive of the idea to handle
the parameter dependence in Eq.~(\ref{Delab})
in the spirit of gauge fixing in massive theories.

\noindent
In the  $b\to \infty$ limit one finds
\begin{equation}
\Pi(a,\infty)=\frac{-{\mbox{\bf P}}^{\left(\frac{3}{2}\right)}+
\frac{p^{2}-m^{2}}{m^{2}}{\mbox{\bf P}}%
_{11}^{\left(\frac{1}{2}\right)}-
a\frac{p^{2}-m^{2}}{p^{2}-am^{2}}
{\mbox{\bf P}}_{22}^{\left(\frac{1}{2}\right)}}%
{p^{2}-m^{2}+i\varepsilon}, \label{propa32}%
\end{equation}
whereas 
for $a\to \infty$ one obtains the propagator that takes
into account the $\gamma\cdot \psi =0$ constraint alone, 
\begin{equation}
\Pi(\infty,b)=\frac{\Delta(\infty, b)}{p^{2}-m^{2}+i\varepsilon} \, .
\end{equation}
Here, 
\begin{eqnarray}
\Delta(\infty,b)&=&{-\mbox{\bf P}}^{\left(\frac{3}{2}\right)}-
\frac{b(p^{2}-m^{2})}{(3\mu^{2}-bm^{2})
(\mu^{2}-bm^{2})-3\mu^{4}} \nonumber \\
&&\left[ \left(  \mu^{2}-bm^{2}\right) {\mbox{\bf P}}_{11}^{\left(\frac{1}{2}\right)}
+\left(
3\mu^{2}-bm^{2}\right)  {\mbox{\bf P}}_{22}^{\left(\frac{1}{2}\right)}-
\sqrt{3}\mu^{2}({\mbox{\bf P}}_{12}^{\left( \frac{1}{2}\right)}
+{\mbox{\bf P}}_{21}^{\left( 
\frac{1}{2}
\right)  }
)\right].
\end{eqnarray}
Notice that neither  $\Pi(a,\infty)$,
nor $\Pi(\infty,b)$ are free from singularities in
the massless limit. 
Nonetheless, the general propagator in Eq.~(\ref{propab})  
that incorporates both  symmetries of the massless theory 
is \underline{not singular} for $m=0$ in which case one recovers 
the propagator in Eq.~(\ref{propabm0}).

In the massive case, the simplest choice for the mass scale 
would be $\mu^{2}=m^{2}$ in which case the general propagator is 
given by Eq.~(\ref{propab}) with the 
$\Delta (a,b)$ operator defined in Eq.~(\ref{Delab}) being replaced by
\begin{eqnarray}
\Delta(a,b)&=&-{\mbox{\bf P}}^{\left(\frac{3}{2}\right)}-\frac{b}{m^{2}}
\frac{p^{2}-m^{2}}{(3-b)(b p^{2}
+a(1-b)m^{2})-3am^{2}} \times \nonumber \\
&& \left[  \left(  b p^{2}+a(1-b)m^{2}\right){\mbox{\bf P}}_{11}
^{\left(\frac{1}{2}\right)}
+a\left(  3-b \right)  m^{2} {\mbox{\bf P}}_{22}
^{\left(\frac{1}{2}\right)}
-\sqrt{3}am^{2}({\mbox{\bf P}}_{12}^{\left( \frac{1}{2}\right) }+
{\mbox{\bf P}}_{21}^{\left( \frac{1}{2}\right)})\right] \, .  \label{Delab_1} %
\end{eqnarray}
 Obviously, this expression is not suited for
taking the $m\to 0$ limit.
Finally, the counterpart to the spin-$1$ Landau propagator is
obtained for  $a=b=0$ in which case only
spin-$\frac{3}{2}$ is propagated,

\begin{equation}
\Pi(0,0)=\frac{-{\mbox{\bf P}}^{\left(\frac{3}{2}\right)}}{p^{2} 
-m^{2}+i\varepsilon}\, .\label{prop00}
\end{equation}

Summarizing this section, in the massless case our equation of motion 
is unique and has as two important symmetries: i) the invariance under 
the gauge transformations in Eq.~(\ref{Gl_1}), and ii) the invariance under 
the point transformations in Eq.~(\ref{point2}).
As long as Lagrangians differing by "rotations" within the 
spin-$\frac{1}{2}$ sector are equivalent,
the massless formalism is unique.  Mass terms break the above symmetries 
in such a way that the  $\gamma\cdot\psi  =0$, and
$\partial\cdot \psi  =0$ conditions (occasionally termed to
as ``gauge'' conditions)  evolve to constraints. 
When properly taken into account, the symmetries
related to these constraints
yield a family of propagators whose spin-$\frac{ 1}{2}$ sectors 
depend on two parameters (termed  to by us  as ``gauge''
parameters in reference to the associated symmetries
in the massless case). 
The propagator in Eq.~(\ref{3/2_PRJCT}) 
represents just one of the members of this family.
In analogy to massive gauge theories, the parameter dependent terms 
in our off-shell propagator can be thought of as terms 
associated with ``gauge fixing''.
Alternatively, the mass terms may be generated via the Higgs mechanism, 
an interesting possibility presently under investigation. 
{}From that perspective, the formalism presented here seems to be a 
good candidate
for the  description of massive spin-$\frac{ 3}{2}$ gauge fields.
Before closing this section we would like to remark that in the conventional 
Rarita-Schwinger formalism it is not possible to interpret the 
$A$-dependence within the context of gauge fixing because 
the invariance under the transformation in Eq.~(\ref{super})
is not general but an exclusive privilege of  the $A=-1$--case
in Eq.~(\ref{Gl_1}), and the symmetry in Eq.~(\ref{point2})
is even completely absent.

%%%%%%%%%%%%%%%%%%%%%%%%%%%%%%%%%%%%%

\section{Interacting spin-$\frac{3}{2}$ particles.}

The interacting theory is now obtained in the standard way
in gauging the Lagrangian in Eq.~(\ref{KNlag}) with the result
\begin{equation}
\mathcal{L}=-(D^{\dagger\mu}\overline{\psi}^{\alpha})\Gamma_{\alpha\beta\mu\nu
}D^{\nu}\psi^{\beta}+m^{2}\overline{\psi}^{\alpha}\psi_{\alpha},
\label{lint}%
\end{equation}
where $D^{\mu}=\partial^{\mu}-ieA^{\mu}$ is the covariant derivative, $(-e)$
denotes the charge of the particle.

This  Lagrangian can be \ written as
\begin{equation}
\mathcal{L}=\mathcal{L}_{free}+\mathcal{L}_{int}\, ,
\end{equation}
with
\begin{eqnarray}
\mathcal{L}_{int}  &=&ie[(\partial^{\nu}\overline{\psi}^{\alpha}%
)\Gamma_{\alpha\beta\nu\mu}\psi^{\beta}-\overline{\psi}^{\alpha}\Gamma_{\alpha\beta\mu\nu
}\partial^{\nu}\psi^{\beta}]A^{\mu}
+e^{2}\overline{\psi}^{\alpha
}\Gamma_{\alpha\beta\mu\nu}\psi^{\beta}A^{\mu}A^{\nu} \nonumber\\
 &=&ej_{\mu}A^{\mu}-e^{2}\overline{\psi}^{\alpha}\Gamma_{\alpha\beta\mu\nu
}\psi^{\beta}A^{\mu}A^{\nu}.
\end{eqnarray}
From the electromagnetic vertex in this Lagrangian we obtain the
electromagnetic transition current in momentum space as%
\begin{equation}
j_{\mu}(p^{\prime},p)=\overline{u}^{\alpha}(p^{\prime})
[-\Gamma_{\alpha\beta\nu\mu}p^{\prime\nu}-\Gamma_{\alpha\beta
\mu\nu}p^{\nu}]u^{\beta}(p),
\end{equation}
where we wrote the free-particle spinors as 
$\psi^{\beta}(x)=u^{\beta}(p)e^{-ip\cdot x}$. 
In order to perform the analogous to the
Gordon decomposition for spin-$\frac{1}{2}$ we write this current in terms of
the four-momentum transfer, $q$, and the summed up four-momenta, $k$,
\begin{equation}
q=p^{\prime}-p,\qquad k=p^{\prime}+p\,,
\end{equation}
to obtain%
\begin{equation}
j_{\mu}(p^{\prime},p)=\overline{u}^{\alpha}(p^{\prime})[
-\Gamma^{S}_{\alpha\beta\mu\nu} k^{\nu}
+\Gamma^{A}_{\alpha\beta\mu\nu} q^{\nu}]u^{\beta}(p),
\label{Noether_Curr}
\end{equation}
where $\Gamma^{S}_{\alpha\beta\mu\nu}$, and $\Gamma^{A}_{\alpha\beta\mu\nu}$ 
stand for the symmetric and anti-symmetric parts under 
the  $\mu\leftrightarrow\nu$ interchanging, respectively,
\begin{equation}
\Gamma^{S}_{\alpha\beta\mu\nu}=\frac{1}{2}(\Gamma_{\alpha\beta\mu\nu}
+\Gamma_{\alpha\beta\nu\mu}),  \qquad
\Gamma^{A}_{\alpha\beta\mu\nu}=\frac{1}{2}(\Gamma_{\alpha\beta\mu\nu}
-\Gamma_{\alpha\beta\nu\mu}).
\end{equation}
It is worth to remark that as long as the tensor 
$\Gamma_{\alpha\beta\mu\nu}$ is 
contracted with the symmetric term $p^{\mu}p^{\nu}$ in the free 
equation of motion (\ref{concise},\ref{Platinoten}), 
only the symmetric part of this tensor is uniquely determined
by the Poincar\'e projector. 
In contrast to this, the antisymmetric part remains ambiguous.
This insight is crucial for the interacting theory since 
it is precisely that very  anti-symmetric part 
that provides essential contributions to the
electromagnetic couplings. 
As a first step in the elucidation of the electromagnetic 
interactions of an elementary spin-$\frac{ 3}{2}$ particle 
we elaborate the interacting theory for the  tensor in 
Eq.~(\ref{tensor_Gamma}). A straightforward calculation yields
\begin{equation}
\Gamma^{S}_{\alpha\beta\mu\nu}
=g_{\alpha\beta}g_{\mu\nu}
-\frac{2}{3}(g_{\mu\alpha}g_{\nu\beta}
+g_{\mu\beta
}g_{\nu\alpha})
 +\frac{1}{6}[(g_{\mu\alpha}\gamma_{\nu}
+g_{\nu\alpha}\gamma_{\mu}%
)\gamma_{\beta}
+\gamma_{\alpha}(g_{\mu\beta}\gamma_{\nu}+g_{\nu\beta}%
\gamma_{\mu})]
-\frac{1}{3}\gamma_{\alpha}\gamma_{\beta} g_{\mu\nu} , \label{Gamma_Sym}
\end{equation}
\begin{equation}
\Gamma^{A}_{\alpha\beta\mu\nu}  
=\frac{1}{3}[g_{\mu\alpha}g_{\nu\beta}-g_{\mu\beta}g_{\nu\alpha}-\frac{i}
{2}g_{\alpha\beta}\sigma_{\mu\nu}]
=-\frac{i}{3}(M_{\mu\nu})_{\alpha\beta}\,,
\label{Gamma_antisym}
\end{equation}
where $(M_{\mu\nu})_{\alpha\beta}$ stand for the (homogeneous) Lorentz group
generators in the vector-spinor representation as given 
in Eq.~(\ref{gensv}). Now, in a 
perturbative calculation one
can use the constraints for the spin-$\frac{3}{2}$ fields in the $\left(
\frac{3}{2}-\frac{3}{2}-\gamma\right)  $ vertex and obtain the following Gordon
decomposition for the transition current
\begin{equation}
j_{\mu}(p^{\prime},p)=\overline{u}^{\alpha}(p^{\prime})[-g_{\alpha\beta
}(p^{\prime}+p)_{\mu}-\frac{i}{3}(M_{\mu\nu})_{\alpha\beta}q^{\nu}
+\frac{2}%
{3}(g_{\mu\alpha}p_{\beta}^{\prime}+g_{\mu\beta}p_{\alpha})]u^{\beta}(p).
\label{Gordon}%
\end{equation}
The latter result exhibits in a transparent way that the 
field described by the gauged equation 
\begin{equation}
\lbrack\Gamma_{\alpha\beta\mu\nu}\pi^{\mu}\pi^{\nu}-m^{2}~g_{\alpha\beta}%
]\psi^{\beta}=0, \label{Platinoint}%
\end{equation}
with the tensor in Eq.(\ref{tensor_Gamma}), carries a gyromagnetic factor of 
$g_{\frac{3}{2}}=\frac{1}{3}$. In addition, the  wave fronts of this gauged equation 
when analyzed along the lines of Refs.~\cite{VZ1,VZ2} 
are found to propagate non-causally.
These findings may look unsatisfactory, indeed,
but as we shall see below they are not to remain the last word
neither on the gyromagnetic ratio, nor on the causality issue.
We shall make the point that causal propagation and gyromagnetic 
ratio are interconnected and that causality requires $g_{\frac{3}{2}}=2$.
The main culprit for the severe underestimation of the gyromagnetic ratio
by Eq.~(\ref{Gordon}) is the incomplete anti-symmetric part of the 
$\Gamma_{\alpha\beta\mu\nu}$ tensor as provided by the space-time invariants.
The correct value of the gyromagnetic ratio is fixed by 
Weinberg's theorem which states that a well behaved 
forward Compton scattering amplitude for a non-strongly interacting
particle with spin $s>\frac{1}{2}$ requires its gyromagnetic factor to
equal $g_{s}=2$ \cite{Weinberg}. The
particular case of the $W$-boson is instructive in that regard because this
particle satisfies Weinberg's principle. Indeed, while the Standard Model
predicts for the $W$-boson $g_{s}=2$, the naive U(1)$_{em}$ gauging of Proca's
equation yields $g_{s}=1$. The difference between these two values is
accounted for by additional contributions coming from the full non-Abelian
$SU(2)_{I}\otimes U(1)_{Y}$ gauge structure in combination with the
spontaneous breaking of the electroweak gauge symmetry. 
On the other hand, more recently, it was also shown that the tree-level 
value of the gyromagnetic ratio of the $\rho^+$-meson is fixed to 2 by 
self-consistency of the corresponding effective quantum field theory 
\cite{Djukanovic}.

Below we shall show how to take advantage of the ambiguities of
the anti-symmetric part of the $\Gamma_{\alpha\beta\mu\nu}$ tensor
and construct a Lagrangian and associated wave equation such that
\begin{itemize}
\item the spin-$\frac{3}{2}$ particle 
is coupled to the electromagnetic field by a gyromagnetic factor
of $g_{\frac{3}{2}}=2$,  
\item  the wave fronts of the solutions of the gauged
equation propagate causally.
\end{itemize}

\subsection{Gauged spin-$\frac{3}{2}$ equation.}

To begin with we first notice that 
the most general anti-symmetric tensor allowed by Lorentz covariance
is given by
\begin{equation}
\Gamma_{\alpha\beta\mu\nu}^{A}=-i\left[  g
\frac{\sigma_{\mu\nu}}{2}g_{\alpha\beta}+ig^{\prime}(g_{\alpha\mu}g_{\beta\nu}-
g_{\alpha\nu}g_{\beta\mu})\right]
+ic(g_{\alpha\mu}\sigma_{\beta\nu}-g_{\alpha\nu}\sigma_{\beta\mu}%
)+id(\sigma_{\alpha\mu}g_{\beta\nu}-\sigma_{\alpha\nu}g_{\beta\mu
})
+if\ \varepsilon_{\alpha\beta\mu\nu}\gamma^{5},
\label{Gamma_AS}
\end{equation}
where $g$, $g^\prime$, $c$, $d$, and $f$ are arbitrary parameters. 
As a consequence, there exist infinitely many equivalent free particle 
theories differing by the values of the above parameters.
However, upon gauging, all these equivalent free particle descriptions 
will become distinguishable through the different values of the multipole
couplings of the spin-$\frac{3}{2}$ particle to the photon field.
Only one of those coupled theories will correspond to the physical reality.
The covariant projector in Eq.~(\ref{Platinoint}) with 
$\Gamma_{\alpha\beta\mu\nu}$ from Eq.~(\ref{tensor_Gamma}) hits the
particular parameter set $g=g^\prime =\frac{1}{3}$, and $c=d=f=0$
which according to our analysis, fails both in the description of
the gyromagnetic ratio as dictated by Weinberg's theorem and in providing 
causal propagation.
We shall remove this shortcoming in 
choosing an appropriate  $\Gamma^{A}_{\alpha\beta\mu\nu}$
that  {\it  ensures causal propagation
of the wave fronts of the solutions of Eq.~(\ref{Platinoint}) within an
electromagnetic environment}. 

In the following we shall first assume $f=0$ for 
simplicity. Then we notice that hermiticity requires $c=d$.

Back to the symmetric $\Gamma_{\alpha\beta\mu\nu}^{S}$ tensor, we observe
that the indices $\alpha $ and $\beta$ have to be moved to the very left,
and the very right, respectively, in order to work in 
 $\pi\cdot \psi $, and $\gamma\cdot \psi$ into the wave equation.
In so doing, one finds various terms in 
$\Gamma_{\alpha\beta\mu\nu}^{S}$  that contain the electromagnetic tensor
according to
\begin{equation}
\Gamma_{\alpha\beta\mu\nu}^{S}\pi^{\mu}\pi^{\nu}
=\pi^{2}g_{\alpha\beta}
+\frac{1}{3}\left(  \gamma_{\alpha}\not \pi-4\pi_{\alpha}\right)  \pi_{\beta}
+\frac{1}{3}(\pi_{\alpha}\not \pi -\gamma_{\alpha}\pi^{2})\gamma_{\beta}     
+\frac{2}{3}ieF_{\alpha\beta}
+\frac{ie}{6}\gamma_{\alpha}\gamma^{\mu}F_{\beta\mu}+
\frac{ie}{6}\gamma^{\mu}F_{\mu\alpha}\gamma_{\beta}\, .
\label{GS_F}
\end{equation}
The latter equation shows that the symmetric part of the
$\Gamma^{S}_{\alpha\beta\mu\nu}$ tensor provides a magnetic
dipole coupling solely for the vector piece of $\psi_{\mu}$
via the $\frac{2}{3}ie F_{\alpha\beta}$ term and leaves the
coupling of the fermionic piece in $\psi_{\mu}$ unspecified.
The latter  comes from the $g$ term in $\Gamma^{A}_{\alpha\beta\mu\nu}$.
In order to see this we cast 
$\Gamma^{A}_{\alpha\beta\mu\nu}\pi^{\mu}\pi^{\nu}$
into the form
\begin{equation}
\Gamma_{\alpha\beta\mu\nu}^{A}\pi^{\mu}\pi^{\nu}  =
-i\left[  g\frac{\sigma
_{\mu\nu}\pi^{\mu}\pi^{\nu}}{2}g_{\alpha\beta}-eg^{\prime}F_{\alpha\beta
}\right]  -2ie c F_{\alpha\beta}
+c(\pi_{\alpha}\not \pi-\not \pi\pi_{\alpha})\gamma_{\beta}+c \gamma_{\alpha
}(\not \pi\pi_{\beta}-\pi_{\beta}\not \pi)\, .
\label{R_A}
\end{equation}
Putting all together results in the following gauged equation:
\begin{eqnarray}%
{\Big(}( \pi^{2}-m^{2})  g_{\alpha\beta}&-&i\left[  g\frac
{\sigma_{\mu\nu}\pi^{\mu}\pi^{\nu}}{2}g_{\alpha\beta}-e
\left(g^{\prime}-2c +\frac{2}{3}\right)%
F_{\alpha\beta}\right]
  +\frac{1}{3}\left(  \gamma_{\alpha}\not \pi
-4\pi_{\alpha}\right)  \pi_{\beta}
+\frac{1}{3}(\pi_{\alpha}\not \pi-\gamma_{\alpha}\pi^{2})\gamma_{\beta}
\nonumber\\
&+&ie\left( \frac{1}{6}-c\right) \gamma^{\mu}F_{\mu\alpha})\gamma_{\beta}
+ie\left( \frac{1}{6}-c\right)
\gamma_{\alpha}\gamma^{\mu}F_{\beta\mu}{\Big)}  \psi^{\beta}=0.
\label{Ec1}%
\end{eqnarray}
The next physical consideration allows to fix 
the $c$-parameter in $\Gamma_{\alpha\beta\mu\nu}^{A}$ and refers to 
the suppression of the $\frac{3}{2}\leftrightarrow \frac{1}{2}$ transitions 
$\gamma\cdot\psi \leftrightarrow \psi^{\mu}$ 
(as required by perturbative calculations)
through  nullifying the  $ie\gamma^{\mu}F_{\mu\alpha}\gamma_{\beta}$-, 
and  $\gamma_{\alpha}\gamma^{\mu}F_{\beta\mu}$ terms, respectively. 
In this manner $c$ is fixed to $c=\frac{1}{6}$ and one is left with the 
two parameters $g$ and $g^\prime$
which in turn describe the gyromagnetic ratios of the fermion-
and vector part of $\psi^{\beta}$. As long as one wishes to have
a spin-$\frac{3}{2}$ coupled by a gyromagnetic factor of
$g_{\frac{3}{2}}$ and given by the Lagrangian
\begin{eqnarray}
{\mathcal L}_{
\mbox{\footnotesize mag}}
&\equiv &-\frac{eg_{\frac{3}{2}}}{2}
\bar \psi^{\alpha} 
( M_{\mu \nu})_{\alpha \beta  }
\psi^{\beta } 
F^{\mu\nu }
=-\frac{e g_{\frac{3}{2}}}{2}\bar\psi^{\alpha } \left(
i(g_{\mu\alpha}
g_{\nu\beta}
-g_{\mu\beta}g_{\nu\alpha})
+\frac{\sigma_{\mu\nu}}{2}g_{\alpha\beta} \right)
F^{\mu\nu}\psi^{\beta }
\nonumber\\
&=&ig_{\frac{3}{2}} \bar\psi^{\alpha}
\left(  \frac{\sigma
_{\mu\nu}\pi^{\mu}\pi^{\nu}}{2}g_{\alpha\beta}-e~F_{\alpha\beta}
\right) \psi^{\beta}\, ,
\label{mag_Lagr}
\end{eqnarray}
one sees that one needs equality of the gyromagnetic ratios
$g_{\frac{1}{2}}$, and $g_{1}$,
in the respective fermion and vector sectors according to
$g_{1}=g_{\frac{1}{2}}=g_{\frac{3}{2}}$.
We here made use of Eq.~(\ref{Lor_gen}) from the Appendix and the relation
$i\sigma_{\mu\nu}=g_{\mu\nu}-\gamma_{\mu}\gamma_{\nu}$.
With this in mind, from now onward we shall assume
$g=g^\prime +\frac{1}{3}\equiv g_{\frac{3}{2}}$, and consider
an interacting  spin-$\frac{3}{2}$ particle described by following 
one-parameter equation
\begin{equation}
\left[  \left(  \pi^{2}-m^{2}\right)  g_{\alpha\beta}-
ig_{\frac{3}{2}}\left(  \frac {
\sigma_{\mu\nu}
\pi^{\mu}
\pi^{\nu} }{2}
g_{\alpha\beta}-e~F_{\alpha\beta}\right)
 +\frac{1}{3}\left(  \gamma_{\alpha}\not \pi
-4\pi_{\alpha}\right)  \pi_{\beta}
+\frac{1}{3}%
(\pi_{\alpha}\not \pi-\gamma_{\alpha}\pi^{2})\gamma_{\beta}\right]
\psi^{\beta}=0 .
\label{gauged_gc}
\end{equation}
We shall fix the $g_{\frac{3}{2}}$-parameter from the requirement on causality.
Before this, we notice that equations like  (\ref{gauged_gc}) are not 
genuine because neither the 
field component $\psi_0$ nor its time-like momentum, $\pi_0$, ever occur.
This behavior reflects the presence of constraints in
Eq.~(\ref{gauged_gc}).
In order to produce a genuine wave equation, one needs
to obtain first the gauged constraints and back-substitute them into
Eq.~(\ref{gauged_gc}).
In subsequently contracting Eq.~(\ref{gauged_gc}) by $\gamma_{\beta}$,
and $\pi_{\beta}$  one obtains the gauged auxiliary conditions
as 
\begin{equation}
\gamma\cdot\psi=\frac{ie}{6m^{2}}\left(  
3g_{\frac{3}{2}}+2\right)  \left(  F_{\mu\beta
}\gamma^{\mu}+i\gamma^{5}\gamma^{\alpha}\tilde{F}_{\beta\alpha}\right)
\psi^{\beta}
\label{aux1}
\end{equation}
and
\begin{eqnarray}
m^{2}\pi\cdot \psi&=&
{\Big[}
ie\left(1-\frac{g_{\frac{3}{2}}}{2}\right)
(F_{\beta\mu}\pi^{\mu}+\pi^{\mu} F_{\beta\mu}) +
ieg_{\frac{3}{2}}\pi^{\alpha}F_{\alpha\beta}
-e\left( \frac{g_{\frac{3}{2}}}{4}+
\frac{1}{6}
\right)
\gamma^{5}
\lbrack 
\gamma^{\alpha}
\widetilde{F}_{\beta\alpha},\not \pi \rbrack
+ie\left( 
\frac{g_{\frac{3}{2}}}{4}-\frac{1}{6}\right)
\nonumber\\
&&\lbrace 
\gamma^{\alpha} F_{\beta\alpha},\not \pi  
\rbrace
{\Big]}\psi^{\beta}
+ie
\left(
\frac{g_{\frac{3}{2}}}{4}-\frac{1}{6}\right)
\gamma^{\nu}(F_{\nu\mu}\pi^{\mu}+\pi^{\mu}F_{\nu\mu})
\gamma\cdot \psi\, ,
\label{aux2}
\end{eqnarray}
respectively.
The resulting equation is now genuine and the wave fronts of its 
solutions would propagate causally provided, the so called 
characteristic determinant of the matrix that
contains only the highest derivatives when replaced
by $n_{\mu}$, the normal vectors to the 
characteristic surfaces,  nullifies only for real values of $n_{0}$
\footnote{We here follow the calculation patterns
of Refs.~\cite{VZ1} and \cite{VZ2}.}. 
The Velo-Zwanziger problem arises because the
characteristic determinant of the (genuine) Rarita-Schwinger equation
allows for $n_{0}$-roots that can become imaginary for sufficiently strong
electromagnetic fields.

\subsection{Causal propagation and gyromagnetic ratio.}

The expression for the matrix that provides the characteristic determinant,
denoted by $D(n, g_{\frac{3}{2}})$, of Eqs.~(\ref{gauged_gc}) with
the substituted Eqs.~(\ref{aux1}) and (\ref{aux2}) is now obtained as
\begin{eqnarray}
D(n, g_{\frac{3}{2}})=&=& |{\mathcal M}_{\alpha\beta}|\, ,\nonumber\\
\mathcal{M}_{\alpha\beta}&=&n^{2}g_{\alpha\beta}+\frac{1}{3}\left(
\gamma_{\alpha}\not n-4n_{\alpha}\right)  N_{\beta}
+\frac{1}{3}\left(
n_{\alpha}\not n-\gamma_{\alpha}n^{2}\right)  \Gamma_{\beta},\nonumber\\
\Gamma_{\beta}  & =&\frac{ie}{6m^{2}}\left(  3g_{\frac{3}{2}}+2\right)  
\left(  F_{\mu\beta
}\gamma^{\mu}+i\gamma^{5}\gamma^{\mu}\tilde{F}_{\mu\alpha}\right)\, , 
\nonumber\\
N_{\beta}  & =&
\frac{1}{m^{2}}\left(
ie(\frac{5}{3}-\frac{3g_{\frac{3}{2}}}{2})F_{\beta\mu}n^{\mu} -
e\left(\frac{g_{\frac{3}{2}}}{4}+\frac{1}{6} \right)
\gamma^{5}\lbrack \gamma^{\alpha}\widetilde{F}_{\beta\alpha},\not n \rbrack
\right)
+\frac{ie}{m^{2}}
\left( \frac{g_{\frac{3}{2}}}{2}-\frac{1}{3}\right)\gamma^{\nu}
F_{\nu\mu}n^{\mu}\Gamma_{\beta}\, .
\label{chrct_det}
\end{eqnarray}
The covariant form of the characteristic determinant
is now calculated to give 
\begin{eqnarray}
&&D(n,g_{\frac{3}{2}})  = \nonumber \\
&&\left(  n^{2}\right)  ^{12}  \left(  
\left[  n^{2}-k^{2}\left(
\frac{5g_{\frac{3}{2}}-2}{4}\right)  ^{2}
\left(  n\cdot F\right)  ^{2}
+ k^{2}\left(
\frac{3g_{\frac{3}{2}}+2}{4}\right)  ^{2}
\left(  n\cdot\tilde{F}\right)  ^{2} \right]^{2} 
+\frac{k^{2}}{4}
\left(  \frac{3g_{\frac{3}{2}}+2}{4}\right)  ^{2}
\left(  \frac{5g_{\frac{3}{2}}-2}{4}\right)  ^{2}
\left(  F\cdot\tilde
{F}\right)  ^{2}\left(  n^{2}\right)  ^{2}  \right) \nonumber\\
 & \times& 
\left(  \left[  n^{2}+k^{2}\left(  \frac{3g_{\frac{3}{2}}+2}{4}\right)  ^{2}
\left[
\left(  n\cdot\tilde{F}\right)  ^{2}-\left(  n\cdot F\right)  ^{2}\right]
\right]  ^{2}
+\frac{k^{4}}{4}\left(  \frac{3g_{\frac{3}{2}}+2}{4}\right)  ^{2}\left(
F\cdot\tilde{F}\right)  ^{2}\left(  n^{2}\right)  ^{2} \right)\, .
\label{Simon_Det}
\end{eqnarray}
Here, $\left(n\cdot F\right)^{\nu}=n_{\mu}F^{\mu\nu}$, 
$\left( n\cdot\tilde{F}\right)^{\nu}=n_{\mu}\tilde{F}^{\mu\nu
}$, $ F\cdot\tilde{F}=F_{\mu\nu}\tilde{F}^{\mu\nu}$, and
$k=\frac{2e}{3m^{2}}$. 
It is quite instructive to compare Eq.~(\ref{Simon_Det})
to the characteristic determinant of the Rarita-Schwinger
equation reported in Ref.~\cite{VZ1} as,
\begin{equation}
D(n)=\left( n^{2} \right)^{4}
\left[ n^{2} +k^{2}\left( \widetilde{F}\cdot n\right)^{2} \right]^{4}\ .
\label{char_det_RS}
\end{equation}
The advantage of Eq.~(\ref{Simon_Det}) over Eq.~(\ref{char_det_RS})
is that in the former case it is possible to factorize
$\left( n^{2}\right)^{16}$ in the expression in the brackets on the cost of
fixing $g_{\frac{3}{2}}$ to either $0$ or $2$, while in the latter
such results impossible. 
To be specific, using
\begin{equation}
\left( n\cdot \widetilde{F} \right)^{2}- \left( n\cdot F\right)^{2}
=-\frac{1}{2}n^{2} F\cdot F\, ,
\label{Hilfs_bez}
\end{equation}
and for $g_{\frac{3}{2}}=0,2$ the characteristic determinant 
takes the following factorized form
\begin{equation}
D(n,g_{\frac{3}{2}}=0,2)=\left(  n^{2}\right)  ^{16}
\left( \left(  1-2k^{2}F\cdot F\right)  ^{2} +
\left( 2k^{2}F\cdot \widetilde{F}\right)^{2}\right)^{2}.
\label{char_det}
\end{equation}
Thus for $g_{\frac{3}{2}}=0,2$, the determinant nullifies only for 
real and field independent $n_0$-values given by
\begin{equation}
n_0=\pm \sqrt{{\mathbf n^{2}}}\, ,
\label{raices}
\end{equation}
and of multiplicity $16$ each, yielding causal propagation. We here 
discard the vanishing gyromagnetic ratio as unphysical
and keep $g_{\frac{3}{2}}=2$ as the physical value.
Compared to this, only eight, i.e. half, of the roots
of the characteristic RS determinant are
necessarily real and  given by
$n_{0}=\sqrt{{\mathbf n}^{2}}$.
In order to find the other eight roots 
it is first quite useful to write explicitly the  
four-vector $\left( n\cdot\widetilde{ F}\right)^{\nu}$ as
\begin{equation}
\left( n\cdot\widetilde{F}\right)^{\nu}=
\left({\mathbf B}\cdot {\mathbf n}, n_0 {\mathbf B} 
-{\mathbf n}\times {\mathbf E}
 \right)\, .
\label{ndotf_vector}
\end{equation}
Substitution in Eq.~(\ref{char_det_RS})
amounts to the following second order equation for
$n_{0}$:
\begin{equation}
n_{0}^{2}(1 -k^{2}{\mathbf B}^{2})
-{\mathbf n}^{2}+k^{2}{\mathbf B}\cdot {\mathbf n}
-k^{2}( -2n_{0}{\mathbf B}\cdot ({\mathbf n}\times {\mathbf E})
+\left({\mathbf n}\times {\mathbf E})^{2} \right) =0\, .
\label{raices_RS}
\end{equation}
As long as the discriminant of the latter equation is 
frame dependent, there are frames where it can become negative and
the roots imaginary.
The frame dependence of Eq.~(\ref{raices_RS}) also shows up
in the possibility of finding frames where the signal velocity
is superluminal, a problem first addressed by Velo and Zwanziger in 
Refs.~\cite{VZ1,VZ2}.

\begin{quote}
The decisive advantage of Eq.~(\ref{gauged_gc}) over the gauged
Rarita-Schwinger equation  is the field-- and
therefore frame-independence of the $n_{0}$-roots, a behavior
which allows for hyperbolicity of the wave equation, causal
signal  propagation, and a gyromagnetic ratio in accord with
the requirements of unitarity in the ultra-relativistic limit.
\end{quote}

The  final form of the $\Gamma_{\alpha\beta\mu\nu}$ tensor now reads
\begin{equation}
\Gamma_{\alpha\beta\mu\nu}=
\Gamma^{S}_{\alpha\beta\mu\nu}
-i{\Big(}
\sigma_{\mu\nu}g_{\alpha\beta}
+i\frac{5}{3}(g_{\alpha\mu}g_{\beta\nu}-g_{\alpha\nu}g_{\beta\mu}){\Big)}
+i\frac{1}{6}(g_{\alpha\mu}\sigma_{\beta\nu}-g_{\alpha\nu}\sigma_{\beta\mu}%
)-i\frac{1}{6}(\sigma_{\alpha\mu}g_{\beta\nu}-\sigma_{\alpha\nu}g_{\beta\mu
}),
\label{causaltensor}
\end{equation}
with $\Gamma_{\alpha\beta\mu\nu}^{S}$ from Eq.~(\ref{Gamma_Sym}). 
It is important to emphasize  that also this tensor satisfies 
Eq.~(\ref{Eqchida}) meaning that the free theory remains same 
as the one related to Eq. (\ref{tensor_Gamma}). 

Finally, a comment is in place on the hermiticity of the equation 
under discussion.
Notice that upon substituting the gauged auxiliary conditions 
from Eqs.~(\ref{aux1}) and (\ref{aux2}) into Eq.~(\ref{gauged_gc})
one does not find a hermitian equation. Above we gave the
causality proof for precisely that very case for the sake of simplicity 
of the expressions and without any loss of generality because
the proof goes through also upon making the equation 
hermitian.

In conclusion, the covariant spin-$\frac{3}{2}$ and mass-$m$ projector
method elaborated here hits the right way toward the consistent
description of spin-$\frac{3}{2}$ within  $\psi_{\mu}$.

\section{Summary.}

In this paper we developed a spin-$\frac{3}{2}$  description on the basis of 
the  Poincar\'e covariant mass-$m$ and spin-$\frac{3}{2} $ 
projectors in $\psi_{\mu}$ and explicitly worked out
the corresponding Lagrangian and wave equation. Our suggested solution to
the problem of the covariant and consistent description of spin-$\frac{3}{2}$
coupled to an electromagnetic field is the fully covariant 
second order equation (\ref{concise},\ref{Platinoten}), 
gauged (\ref{gauged_gc}) and with 
the tensor $\Gamma_{\alpha\beta\mu\nu}$ given in Eq.~(\ref{causaltensor}).
 We studied the symmetries of the suggested Lagrangian in the massless limit
and their extrapolation to the massive case, where they gave rise to
constraints and introduced  parameter dependence of the off-mass--shell
propagators. We observed that the off-shell massive spin-$\frac{3}{2}$ 
propagator suggested by us is of 
the type of propagators that appear in the massive gauge theories. 
{}From that we concluded that its parameter dependence is actually 
brought about by the symmetries of the massless theory, one of them being
the gauge freedom, and as such can be handled by means of ``gauge'' fixing. 
We introduced electromagnetic interactions
into the theory and showed that the wave fronts of the solutions of
the gauged equation propagate causally provided
the gyromagnetic factor of the spin-$\frac{3}{2}$ particle were
to be $g_{\frac{3}{2}}=2$ as required by unitarity in the ultra-relativistic
limit. The structure of the off-mass-shell propagator, and
the  causal propagation in combination with a gyromagnetic
ratio of $g_{\frac{3}{2}}=2$ seem to qualify  
the formalism elaborated here as a promising
candidate for the consistent description of massive spin-$\frac{3}{2}$ 
gauge fields. 

\section*{Acknowledgments.}

Work supported by Consejo Nacional de Ciencia y Tecnologia (CONACyT) Mexico
under projects 37234-E and C01-39820.

\section{Appendix.}

In this appendix we collect conventions and some results on the symmetry of
spacetime under rotations, boost and translations transformations that
constitute the Poincar\'{e} group for which the squared Pauli-Lubanski vector
is a Casimir invariant. In terms of the Poincar\'{e} group generators,
$M_{\mu\nu}$ and $p_{\eta}$ and their algebra \cite{WKT}
\begin{eqnarray}
\left[  M_{\mu\nu},M_{\alpha\beta}\right]   &  =&-i(g_{\mu\alpha}M_{\nu\beta
}-g_{\mu\beta}M_{\nu\alpha}+g_{\nu\beta}M_{\mu\alpha}-g_{\nu\alpha}M_{\mu
\beta})\,,\label{M_comm}\nonumber\\
\left[  M_{\alpha\beta},p_{\mu}\right]   &  =&-i(g_{\mu\alpha}p_{\beta}%
-g_{\mu\beta}p_{\alpha}),\nonumber\\
\left[  p_{\mu},p_{\nu}\right]  &=&0\,,
\label{M_P_comm}%
\end{eqnarray}
where $g_{\mu\nu}$=diag$(1,-1,-1,-1)$ is the metric tensor, the
Pauli--Lubanski (PL) vector is defined as
\begin{equation}
\mathcal{W}_{\mu}=\frac{1}{2}\epsilon_{\mu\nu\alpha\beta}M^{\nu\alpha}%
p^{\beta}\,, \label{paulu}%
\end{equation}
with $\epsilon_{0123}=1$. This operator can be shown to satisfy the following
commutation relations
\begin{eqnarray}
\lbrack M_{\mu\nu},{}\mathcal{W}_{\alpha}]&=&-i(g_{\alpha\mu}{}\mathcal{W}_{\nu
}-g_{\alpha\nu}{}\mathcal{W}_{\mu}),\quad\lbrack\mathcal{W}_{\alpha},p_{\mu
}]=0,\nonumber\\
\lbrack\mathcal{W}_{\alpha},\mathcal{W}_{\beta}]&=&-i\epsilon
_{\alpha\beta\mu\nu}{}\mathcal{W}^{\mu}p^{\nu}\,, \label{conmrelpl}%
\end{eqnarray}
i.e. it transforms as a four-vector under Lorentz transformations. Moreover,
its square commutes with all the generators and is a group invariant. For this
reason elementary particles are required to transform invariantly under the
action of ${}\mathcal{W}^{2}$ and to be labeled by the ${}\mathcal{W}^{2}$
eigenvalues next to those of $p^{2}$.

In general, for a specific representation, the generators of the (homogeneous)
Lorentz group $M_{\mu\nu}$ carry additional indices. We denote these indices
by capital Latin letters as ($M_{\mu\nu})_{AB}$. The Lorentz group generators
in the vector (i.e.$\left(  \frac{1}{2},\frac{1}{2}\right)  $) and spinor
(i.e. $(\frac{1}{2},0)\oplus(0,\frac{1}{2})$ ) space are respectively given
as
\begin{equation}
\left(  M_{V}^{\rho\sigma}\right)  _{\alpha\beta}=i(g_{~\alpha}^{\rho
}g_{~\beta}^{\sigma}-g_{~\beta}^{\rho}g_{~\alpha}^{\sigma}),\qquad(M_{S}%
^{\rho\sigma})_{ab}=\frac{1}{2}(\sigma^{\rho\sigma})_{ab}. \label{Lor_gen}%
\end{equation}
The indices $A,B$ are Lorentz indices (in the vector basis) for the vector
space: $A=\{\alpha\},B=\{\beta\}$, whereas for the spinor representation they
are spinorial indices: $A=\{a\},B=\{b\}$. {}The Pauli-Lubanski operators in
vector and spinor space, denoted respectively by $W^{\lambda}$, $w^{\lambda}$
have the explicit form
\begin{equation}
\left[  W_{\lambda}\right]  _{\alpha\beta}=i\epsilon_{\lambda\alpha\beta\mu
}p^{\mu},\qquad(w_{\lambda})_{ab}=\frac{i}{2}(\gamma_{5}\sigma_{\lambda\nu
})_{ab}p^{\nu}\,. \label{PL_VD}%
\end{equation}
The squared Pauli- Lubanski operators are now calculated as
\begin{eqnarray}
\left[  W^{2}\right]  _{\alpha}^{\quad\beta}  &  =&-2\left(  g_{\alpha\beta
}g_{\mu\nu}-g_{\alpha\nu}g_{\beta\mu}\right)  p^{\mu}p^{\nu}\, ,
\label{W2_hh}\\
\lbrack w^{2}]_{ab}  &  =&-\frac{1}{4}(\sigma_{\lambda\mu})_{ac}(\sigma
_{\quad\nu}^{\lambda})_{cb}p^{\mu}p^{\nu}\,. \label{W2_Dirac}%
\end{eqnarray}
In particular, our general equation (\ref{prj_eq}) for the spin-$1$ subspace
in $\left(  \frac{1}{2},\frac{1}{2}\right)  $ reads%
\begin{equation}
\left[  W^{2}\right]  _{\alpha}^{\quad\beta}A_{\beta}=-2m^{2}A_{\alpha}\,.
\label{Proca_W2}%
\end{equation}
As for the vector-spinor space, the (homogeneous) Lorentz group generators are
given by
\begin{equation}
(M^{\rho\sigma})_{\alpha\beta ab}=\left(  M_{V}^{\rho\sigma}\right)
_{\alpha\beta}\delta_{ab}+g_{\alpha\beta}(M_{S}^{\rho\sigma})_{ab}.
\label{gensv}%
\end{equation}
The Pauli-Lubanski vector for the vector-spinor representation reads:
\begin{eqnarray}
(\mathcal{W}^{\lambda})_{\alpha a\beta b}=(W^{\lambda})_{\alpha\beta}%
\delta_{ab}
&+&g_{\alpha\beta}(w^{\lambda})_{ab}\,. \label{gen_dir_prdct}%
\end{eqnarray}
The indices $A,B$ in this case correspond to the sets $A=\{\alpha
a\},B=\{\beta b\}$. The squared Pauli-Lubanski operator in the vector-spinor
representation reads
\begin{equation}
(\mathcal{W}^{2})_{\alpha\beta ab}=(W^{2})_{\alpha\beta}\delta_{ab}%
+(W)_{\alpha\beta}\cdot(w)_{ab}+(w)_{ab}\cdot(W)_{\alpha\beta}
+g_{\alpha\beta}(w^{2})_{ab}. \label{PL_VS}%
\end{equation}
We obtain the involved operators as%
\begin{eqnarray}
(W^{2})_{\alpha\beta}  &  =&-2\left(  g_{\alpha\beta}g_{\mu\nu}-g_{\alpha\nu
}g_{\beta\mu}\right)  p^{\mu}p^{\nu},\qquad
(w^{2})_{ab} =-\frac{1}{4}%
(\sigma_{\lambda\mu}\sigma_{\quad\nu}^{\lambda})_{ab}\ p^{\mu}p^{\nu},
\nonumber\\
(W\cdot w+w\cdot W)_{\alpha a\beta b}   & =& -\frac{1}{2}(\epsilon_{\quad
\alpha\beta\mu}^{\lambda}\gamma^{5}(\sigma_{\lambda\nu})_{ab}
+\epsilon
_{\quad\alpha\beta\nu}^{\lambda}\gamma^{5}(\sigma_{\lambda\mu})_{ab})\ p^{\mu
}p^{\nu}\,. \label{W2_VD_VS}%
\end{eqnarray}
In substituting Eqs.~(\ref{W2_VD_VS}) into (\ref{PL_VS}) results in
\begin{equation}
(\mathcal{W}_{\lambda}\mathcal{W}^{\lambda})_{\alpha\beta ab}=T_{\alpha a\beta
b\mu\nu}p^{\mu}p^{\nu}\, ,
\end{equation}
with%
\begin{equation}
T_{\alpha a\beta b\mu\nu}=-2\left(  g_{\alpha\beta}g_{\mu\nu}-g_{\alpha\nu
}g_{\beta\mu}\right)  \delta_{ab}-\frac{1}{4}g_{\alpha\beta}(\sigma
_{\lambda\mu}\sigma_{\quad\nu}^{\lambda})_{ab}
-\frac{1}{2}(\epsilon
_{\quad\alpha\beta\mu}^{\lambda}\gamma^{5}(\sigma_{\lambda\nu})_{ab}
+\epsilon_{\quad\alpha\beta\nu}^{\lambda}\gamma^{5}(\sigma_{\lambda\mu}%
)_{ab}). \label{tensor}%
\end{equation}

\end{document}